\documentclass[12pt,onecolumn,conference,compsoc]{IEEEtran}
\usepackage[cmex10]{amsmath}
\usepackage{array}
\usepackage{mdwmath}

\usepackage{tikz}

\usetikzlibrary{arrows,%
                shapes,positioning}

\usepackage{amssymb}
\usepackage{graphicx}
\advance \oddsidemargin by 0.5in
\newcommand{\myparskip}{1pt}
\parskip \myparskip

\newtheorem{theorem}{Theorem}
\newtheorem{cor}[theorem]{Corollary}
\newtheorem{lemma}[theorem]{Lemma}
\newtheorem{prop}[theorem]{Proposition}

\newtheorem{conj}[theorem]{Conjecture}

\begin{document}
\title{On the Relationships among Optimal Symmetric Fix-Free Codes}
\author{S. M. Hossein Tabatabaei Yazdi \\ 
Texas A\&M University \\ smhtyazdi@neo.tamu.edu 
\and Serap A.~Savari \\ 
Texas A\&M University \\ savari@ece.tamu.edu}
\maketitle{}
\vspace*{-0.2in}
\begin{abstract}
\vspace*{-0.1in}
Symmetric fix-free codes are prefix condition codes in which each codeword is
required to be a palindrome.  Their study is motivated by the topic of joint
source-channel coding.  Although they have been considered by a few 
communities they are not well understood.  In earlier work we used a 
collection of instances of Boolean satisfiability problems as a tool in the 
generation of all optimal binary symmetric fix-free codes with $n$ codewords 
and observed that the number of different optimal codelength sequences grows
slowly compared with the corresponding number for prefix condition codes.  
We demonstrate that all optimal symmetric fix-free codes can 
alternatively be obtained by sequences of codes 
generated by simple manipulations starting from one particular code.
We also discuss simplifications in the process of searching for this set of
codes.
\end{abstract}
\vspace*{-0.2in}
\section{Introduction}
\vspace*{-0.1in}
Shannon's
pioneering work on information theory \cite{key-shannon} establishes that 
source and channel encoding can be separated without a loss of performance
assuming infinite blocklengths are permitted.  However, that result does not
apply to real transmission situations with complexity and latency constraints,
and there is therefore an interest in joint source-channel coding and decoding
techniques.  
Many video, audio, and image standards use prefix condition codes.
It is therefore interesting to devise prefix condition codes with additional
constraints which result in binary encodings of data with increased immunity
to noise prior to channel encoding.  For example, {\em fix-free} or
reversible variable length codes (see, e.g., \cite{mps, gm, bp, twm}) are
prefix condition codes in which no codeword is the suffix of another codeword,
and they are components of the video standards H.264 and MPEG-4
\cite{tw, lv, wsbl, mws}.

Our focus in this paper is upon a subclass of fix-free codes known as
{\em symmetric} fix-free codes \cite{twm}.  Here each codeword must be a
palindrome.  Symmetric fix-free codes were found \cite{bh00} to be preferable
to other fix-free codes for joint source-channel coding.  They are also
easier to study because a collection of palindromes which satisfies the prefix
condition automatically satisfies the suffix condition 
\cite{twm, tsai, savari2009optimal}.  Nevertheless, although they have also
been studied in \cite{bh01, tw, tseng2003construction, aks, hwh, stak}
they are not well-understood.  For example, 
there is no exact counterpart to the Kraft inequality/equality for symmetric
fix-free codes, although \cite{twm, tsai, savari2009optimal, stak} discuss
some simple nonexhaustive necessary and sufficient conditions for the codeword
lengths of such codes.  In \cite{savari2009optimal, aks, stak} we convert
the problem of determining the existence of a symmetric fix-free code with
given codeword lengths into a Boolean satisfiability problem and offer
branch-and-bound algorithms to find the set of optimal codes for all 
memoryless sources, i.e., codes which minimize the average codeword length 
among all symmetric fix-free codes for some choice of source probabilities.
For a given source its optimal code can be found by calculating the expected 
codeword length for each of the optimal codelength sequences and choosing 
the corresponding optimal code.  
In \cite{aks, stak} we show that the number of sorted and nondecreasing 
optimal codelength sequences for binary symmetric fix-free codes with $n$ 
codewords appears to grow very slowly with $n$ compared with the 
corresponding exponential growth \cite{eng} for binary prefix condition codes
(see the appendix).  
Therefore, when $n$ is not too large it appears to be feasible to calculate 
and store all optimal codes and to choose the best among them for a given 
application.  The paper \cite{hwh}
proposes an $A^{*}$-based algorithm for a different way to obtain an optimal
symmetric fix-free code for a given source, but this procedure does not
offer much mathematical insight about optimal codes.  The existing
understanding about optimal codes is very limited.

Although solving instances of Boolean satisfiability problems 
can be one component in the generation of optimal codes, we propose in 
Section~3 a completely different derivation of them.
Our inspiration comes from a paper \cite{pr} which shows
that the space of all sorted and non-decreasing sequences of codeword lengths
of optimal binary prefix condition codes forms a lattice called the 
{\em imbalance} lattice.  Among the length sequences which satisfy the
Kraft inequality with equality, 
$(1, \ 2, \ 3, \ \dots , \ n-1, \ n-1)$ is considered to be the most
imbalanced because it corresponds to the largest sum of codeword lengths.
The authors of \cite{pr} describe a basic operation on three values of a
codeword length sequence which when repeated enough times will transform
the most imbalanced codeword length sequence into 
an arbitrary sorted and non-decreasing optimal codeword length sequence.

We will not work here with length sequences but instead with the binary
codes themselves.  Although the optimal codes do not form a lattice
we will see that they can each be attained from the repetition of a basic
operation which eventually transforms the most ``imbalanced'' optimal code
into an arbitrary optimal code.  (The basic operation here is completely
different from that of \cite{pr}, and the number of codewords
it will affect in one application depends on several factors.)
The following results from \cite{stak} show that the most imbalanced optimal
symmetric fix-free code is $\{ 0 , \ 11, \ 101, \ 1001, \ \dots \}$ with 
length sequence $(1, \ 2, \ \dots , \ n)$.

\begin{prop}
\cite[Prop.~2.2]{stak}
The code $\{ 0 , \ 11, \ 101, \ 1001, \ \dots \}$ with $n \geq 3$ codewords
is in the set of optimal symmetric fix-free codes with $n$ codewords.
\end{prop}
\begin{theorem}
\cite[Thm.~2.5]{stak}
The sorted and non-decreasing length sequence \\ 
$(l_1, \ l_2, \ \dots , \ l_n)$
of an optimal binary symmetric fix-free code with $n$ codewords satisfies
$l_i \leq n$ for $i \in \{1, \ 2, \ \dots , \ n\}$ and
$\sum_{i=1}^n l_i \leq \sum_{i=1}^n i = n (n+1) /2$.
\label{thm:stak}
\end{theorem}

Our initial procedure to generate any optimal symmetric fix-free code 
will also generate some suboptimal codes.
Part of the contribution of Section~4 is to provide simple tests to reduce 
the number of candidates for optimal codes, and one of these tests
can be viewed as a generalization of Theorem~\ref{thm:stak}.
\vspace*{-.1in}
\section{Preliminaries}
\vspace*{-.1in}
Given a palindrome $\sigma$, we define the set of its 
{\em neighboring palindromes} ${\cal N}(\sigma)$ by
\begin{displaymath}
{\cal N}(\sigma) = \{ \mbox{palindromes $w$: $\sigma$ is the longest 
palindrome which is a proper prefix of $w$} \} .
\end{displaymath}
For example, ${\cal N}(0) = \{ 00, \ 010, \ 0110, \ \dots \}$.
For any string $w$, let $|w|$ denote the length of $w$.  We will be
interested in the following (possibly empty) subset of ${\cal N}(\sigma)$
\begin{displaymath}
{\cal N}_n (\sigma) = \{ w \in {\cal N}(\sigma) : \; |w| \leq n \}.
\end{displaymath}
Note that if we remove a palindrome $\sigma$ from a symmetric fix-free
code, then we can add to the remainder of that code any subset of
${\cal N} (\sigma)$ to obtain another symmetric fix-free code with possibly
more codewords than the original code. 

Observe that for any symmetric fix-free code 
$C_n = \{c_1, \ c_2, \ \dots, \ c_n\}$, we can define a ``complementary''
symmetric fix-free code by reversing the bits of each codeword.
For $n \geq 3$ any symmetric fix-free code with have at most one codeword
consisting of a single bit, so we can assume without loss of generality 
that $1 \not\in C_n$.  We will ultimately be concerned with the set
$\mathbb{O}_{n}$ of optimal symmetric fix-free codes $C_n$ with $n$ codewords
for which $1 \not\in C_n$.  However, we begin by considering the larger set
$\mathbb{S}_{n}$ of symmetric fix-free codes $C_n$ with $n$ codewords
for which $1 \not\in C_n$ and $\max_{1 \leq i \leq n} |c_i| \leq n$.

We will call the symmetric fix-free code 
$\{ 0 , \ 11, \ 101, \ 1001, \ \dots \}$ with length sequence
$ (1, \ 2, \ \dots , \ n)$ the {\em root} code of length $n$ and label it
$R_n$.  We have the following result.
\begin{lemma}
Any codeword of a symmetric fix-free code $C_n \in \mathbb{S}_{n}$ has a
codeword of $R_n$ as a prefix.
\label{lem:root}
\end{lemma}
\begin{IEEEproof}
Let $s_i, \ i \leq n,$ denote the codeword of length $i$ in $R_n$.
All codewords in $C_n$ which begin with a $0$ have $s_1$ as the prefix.
All other codewords in $C_n$ begin with a $1$, and by assumption, 
$1 \not\in C_n$.  Observe that any binary string beginning with a $1$ and
having length between $2$ and $n$ will either have $s_i$ as a prefix
for some $2 \leq i \leq n$ or it will be in the set 
$\{10, \ 100, \ 1000, \ \dots\}$.  However, a binary string beginning with
a $1$ and ending with a $0$ is not a palindrome and is therefore not in $C_n$.
\end{IEEEproof}
\vspace*{-.1in}
\section{Relations among Optimal Symmetric Fix-Free Codes}  
\vspace*{-.1in}
We define two relations $\rightarrow$ and $\Rightarrow$ between codes 
$S_n, \ \hat{S}_n \in \mathbb{S}_{n}$ by
\begin{eqnarray*}
S_n \rightarrow \hat{S}_n & & \mbox{if there exists $\sigma \in S_n$
such that} \; \hat{S}_n \subseteq S_n \cup {\cal N}_n (\sigma) \setminus 
\{\sigma \}. \\
& & \mbox{For this $\sigma$ we write} \;
S_{n} \overset{\sigma}{\rightarrow} \hat{S}_n  . \\
S_n \Rightarrow \hat{S}_n & & \mbox{if there exists $\sigma \in S_n$
such that $\hat{S}_n$ consists of the shortest $n$ words of} \\
& &  S_n \cup {\cal N}_n (\sigma) \setminus \{ \sigma \} . \;
\mbox{For this $\sigma$ we write $S_{n} \overset{\sigma}{\Rightarrow} 
\hat{S}_n .$}
\end{eqnarray*}

We have the following result about $\mathbb{S}_{n}$.
\begin{theorem}
For any code $C_n \in \mathbb{S}_{n}$ with codeword lengths
$l_1, \ l_2, \ \dots , \ l_n$, there exists an integer 
$m \leq \sum_{i=1}^n (l_i-1) = O(n^2)$
and a sequence of symmetric fix-free codes 
\mbox{$S_n^{(1)}, \ S_n^{(2)}, \ \dots , \ S_n^{(m)} \in \mathbb{S}_{n}$}
for which 
$R_n = S_n^{(0)} \rightarrow S_n^{(1)} \rightarrow S_n^{(2)} \rightarrow 
\dots \rightarrow S_n^{(m)} = C_n$
and with the property that each codeword of $C_n$ has a prefix in $S_n^{(i)}$
for each $i \in \{0, \ 1, \ \dots, \ m-1 \}$.
Furthermore, there exists a code $B_n \in \mathbb{S}_{n}$ for which the 
preceding sequence requires $m= \Omega (n^{1.5})$ codes. 
\label{thm:rightarrow}
\end{theorem}
\begin{IEEEproof}
Consider the following algorithm to generate the codes
$S_n^{(1)}, \ S_n^{(2)}, \ \dots , \ S_n^{(m)}$:
\begin{enumerate}
\item $S_n^{(0)} = R_n ; \; i=0$.
\item If there exists a codeword $w \in C_n$ which has a proper prefix
$\sigma \in S_n^{(i)}$:
\begin{enumerate}
\item Find the subset $C_n (\sigma)$ of ${\cal N}_n (\sigma)$ consisting of 
the strings which are prefixes of codewords of the code $C_n$.  If there are
$\# (\sigma)$ words in $C_n (\sigma)$, then there is a subset
$D^{(i)} \subset S_n^{(i)} \setminus \{\sigma\}$ with $\# (\sigma) -1$ strings
such that no element of $D^{(i)}$ is a prefix of a word in $C_n$.
\item Set $S_n^{(i+1)} = S_n^{(i)} \cup C_n (\sigma) \setminus
\{ \{ \sigma \} \cup D^{(i)} \}$.
\end{enumerate}
\item $i \leftarrow i+1$.  Goto 2.
\end{enumerate}

We argue inductively that this procedure generates an appropriate sequence of
codes.  For the basis step, we have seen in Lemma~\ref{lem:root} that every
element of $C_n$ has a prefix in $R_n = S_n^{(0)}$.  For the inductive
step, assume that every element of $C_n$ has a prefix in $S_n^{(k)}$ for some
$k \geq 0$, and assume $w \in C_n$ has a proper prefix $\sigma$
in $S_n^{(k)}$.  Since
${\cal N}_n (\sigma)$ contains the palindromes of length at most $n$ for
which $\sigma$ is the longest proper prefix which is a palindrome,
$w$ has a prefix (possibly the full string) which is an element of
${\cal N}_n (\sigma)$.  That prefix will be a member of $S_n^{(k+1)}$,
and we repeat this argument for any other codeword of $C_n$ having 
$\sigma$ as a prefix.
For each codeword of $C_n$ having a different prefix in $S_n^{(k)}$,
we assume that the same prefix will be an element of $S_n^{(k+1)}$.
Therefore $S_n^{(k+1)}$ has the desired property.

For an upper bound on $m$, each application of operation $\rightarrow$ will 
involve a different choice for the string $\sigma$, and each one will be a 
palindrome which is a proper prefix of at least one codeword.  The result 
follows since each codeword of length $l_i, \ i \in \{1, \ 2, \ \dots, \ n\},$ 
has $l_i -1 \leq n-1$ proper prefixes.

For the last part, our code $B_n$ will consist of $n$ palindromes of length
$n$ which begin with and end with $0$.  For convenience we assume here that
$n$ is even.  Since there are $2^{0.5n-1}$ such palindromes, we must have
$n \geq 8$.  We will describe the code in terms of $l$ clusters of codewords.
The first cluster is the all-zero string, which has $n-1$ proper prefixes
all of which are palindromes.  The second cluster is a single string with
left half $0101 \dots$.  The new proper prefixes which are palindromes are
$010, \ 01010, \ \dots$, and there are $({1}/{2}) \cdot (0.5n - 2 - O(1))$ 
of them.  The third cluster consists of the two strings with left half 
$0110110110 \dots$ and left half $00100100 \dots$.  The new proper prefixes
of the left halves of these string
which are palindromes are $0110, \ 00100, \ 0110110, \ 00100100, \ \dots$,
and there are $({2}/{3}) \cdot (0.5n - 3 - O(1))$ of them.  Cluster $j, \
j \in \{2, \ 3, \ \dots , \ l\}$, consists of $j-1$ strings.  The left half
of string $k \in \{1, \ \dots , \ j-1\}$ of cluster $j$ is a repetition
of the length $j$ string beginning with $k$ zeroes and ending with
$j-k$ ones.  There are $((j-1)/{j}) \cdot (0.5n - j - O(1))$ proper prefixes
of the left halves of these strings.  Since there
are $n$ words in the combination of all clusters, we have that 
$l = \Omega (\sqrt{n})$, and the number of proper prefixes of all $n$
codewords is $\Omega (n^{1.5})$.
\end{IEEEproof}

We can characterize the set of optimal codes as follows.
\begin{theorem}
For any code $C_n \in \mathbb{O}_{n}$ there exist an integer $m=O(n^2)$
and a sequence of symmetric fix-free codes  
\mbox{$S_n^{(1)}, \ S_n^{(2)}, \ \dots , \ S_n^{(m)} \in \mathbb{S}_{n}$}
for which
$R_n = S_n^{(0)} \Rightarrow S_n^{(1)} \Rightarrow S_n^{(2)} \Rightarrow 
\dots \Rightarrow S_n^{(m)} = C_n.$
\label{thm:Rightarrow}
\end{theorem}
\begin{IEEEproof}
By Theorem~\ref{thm:rightarrow}, there exist $m=O(n^2)$,
a sequence of codes  
$C_n^{(1)}, \ C_n^{(2)}, \ \dots ,$ $C_n^{(m)} \in \mathbb{S}_{n}$,
and palindromes $w_i \in C_n^{(i)}, \ 0 \leq i \leq m-1,$
such that
$R_n = C_n^{(0)} \overset{w_0}\rightarrow C_n^{(1)} 
\overset{w_1}\rightarrow C_n^{(2)} \overset{w_2}\rightarrow 
\dots \overset{w_{m-1}}\rightarrow C_n^{(m)} = C_n ;$ i.e.,
\begin{equation}
C_n^{(i+1)} \subseteq C_n^{(i)} \cup {\cal N}_n (w_i) 
\setminus \{w_i \} . \label{eq:w}
\end{equation}
Let $k \geq 1$ be the smallest integer for which 
$C_n^{(k-1)} \not\Rightarrow C_n^{(k)}$, and let $S_n^{(k)}$ denote the 
choice of the shortest $n$ strings in
$C_n^{(k-1)} \cup {\cal N}_n (w_{k-1}) \setminus \{w_{k-1} \}$ which has 
maximum overlap with $C_n^{(k)}$. 
Therefore, for any $c \in C_n^{(k)} \setminus S_n^{(k)}$,
\begin{equation}
|c| \geq \max_{s \in S_n^{(k)}} |s|. \label{eq:cs}
\end{equation}
Since by assumption $C_n^{(m)} = C_n \in \mathbb{O}_{n}$, we must have $k<m$.
We will finish the proof by showing that regardless of the value of $k$, there
is a way to effectively increase it by one.  More precisely, we establish the
following result:
\begin{lemma}
For the codes $S_n^{(k)}$ and $C_n$ defined above, there is an integer 
$d \leq m-k$ and codes 
\mbox{${\hat{S}_n}^{(k+1)}, \ {\hat{S}_n}^{(k+2)}, \ \dots , \ 
{\hat{S}_n}^{(k+d)} \in \mathbb{S}_{n}$}
for which
$S_n^{(k)} \rightarrow {\hat{S}_n}^{(k+1)} \rightarrow {\hat{S}_n}^{(k+2)} 
\rightarrow \dots \rightarrow {\hat{S}_n}^{(k+d)} = C_n.$
\end{lemma}
\begin{IEEEproof}
By assumption, $S_n^{(k)} \neq C_n$.  For $i \in \{k, \ k+1, \ \dots , m\}$,
define
\begin{eqnarray}
F^{(i)} & = & \{ \sigma \in C_n^{(i)} : \ \mbox{$\sigma$ has a prefix in
$S_n^{(k)}$} \} \label{eq:f} \\
\mbox{and $G^{(i)}$} & = & 
\{ \sigma \in C_n^{(i)} : \ \mbox{$\sigma$ has no prefix in $S_n^{(k)}$} \} .
\label{eq:g} 
\end{eqnarray}
The sets $F^{(i)}$ and $G^{(i)}$ are clearly disjoint, and
\begin{equation}
C_n^{(i)} = F^{(i)} \cup G^{(i)}. \label{eq:u} 
\end{equation} 
For $i \geq k$, each $w_i$ defined by (\ref{eq:w}) satisfies
$w_i \in F^{(i)}$ or $w_i \in G^{(i)}$, but not both.
Consider the case where $w_i \in G^{(i)}, \; w_i \not\in F^{(i)}$.
By (\ref{eq:w}) and (\ref{eq:u}), 
\begin{equation}
F^{(i+1)} \subseteq 
F^{(i)} \cup ((G^{(i)} \setminus \{w_i\}) \cup {\cal N}_n (w_i) )  . 
\label{eq:c}
\end{equation}
By the argument used in the proof of Theorem~\ref{thm:rightarrow}, 
every element of $G^{(i)}$ has a prefix in $C_n^{(k)}$.  
Therefore the definition of $G^{(i)}$ implies that each of its elements,
including $w_i$, has a prefix in $C_n^{(k)} \setminus S_n^{(k)}$.
Hence every element of the sets ${\cal N}_n (w_i)$ and
$(G^{(i)} \setminus \{w_i \}) \cup {\cal N}_n (w_i) $
has a prefix in $C_n^{(k)} \setminus S_n^{(k)}$.
To arrive at a contradiction, suppose 
$v \in (G^{(i)} \setminus \{w_i \}) \cup {\cal N}_n (w_i)  $
has a prefix in $S_n^{(k)}$, say $s$.  Let $c$ be the prefix of $v$ in
$C_n^{(k)} \setminus S_n^{(k)}$.  Since both $s$ and $c$ are prefixes of $v$,
either $s$ is a prefix of $c$ or $c$ is a prefix of $s$.  Observe that
$s, \ c \in  C_n^{(k)} \cup S_n^{(k)}$, and so $C_n^{(k)} \cup S_n^{(k)}$
does not satisfy the prefix condition.  However, $C_n^{(k)} \cup S_n^{(k)}$
is a symmetric fix-free code because the rules for constructing $C_n^{(k)}$
and $S_n^{(k)}$ imply that 
\begin{displaymath}
C_n^{(k)} \cup S_n^{(k)} \subseteq (C_n^{(k-1)} \setminus \{w_{k-1} \} )
\cup {\cal N}_n (w_{k-1}) , 
\end{displaymath}
and the right-hand side of the preceding relation describes a symmetric
fix-free code.  This contradiction implies that no element of
$(G^{(i)} \setminus \{w_i \}) \cup {\cal N}_n (w_i) $ has a
prefix in $S_n^{(k)}$.  Therefore, we find from (\ref{eq:c}) that
\begin{equation}
F^{(i+1)} \cap ((G^{(i)} \setminus \{w_i \}) \cup {\cal N}_n (w_i) ) 
= \emptyset . \label{eq:i}
\end{equation}
Therefore (\ref{eq:c}) and (\ref{eq:i}) imply that for $i \geq k$,
\begin{equation}
F^{(i+1)} \subseteq F^{(i)} \; \mbox{if $w_i \not\in F^{(i)}$ } . \label{eq:ff}
\end{equation}

In the derivation of (\ref{eq:i}) we argued that $C_n^{(k)} \cup S_n^{(k)}$
is a symmetric fix-free code and hence satisfies the prefix condition.
Observe that 
$C_n^{(k)} \cup S_n^{(k)} = (C_n^{(k)} \setminus S_n^{(k)}) \cup S_n^{(k)}$.
Therefore no element of $C_n^{(k)} \setminus S_n^{(k)}$ has a prefix in
$S_n^{(k)}$, or equivalently,
\begin{equation}
C_n^{(k)} \setminus S_n^{(k)} \subseteq G^{(k)} . \label{eq:csg}
\end{equation}
Since every element of $C_n^{(k)} \cap S_n^{(k)}$ has a prefix in $S_n^{(k)}$,
it follows that
\begin{equation}
C_n^{(k)} \cap S_n^{(k)} \subseteq F^{(k)} . \label{eq:csf}
\end{equation}
By (\ref{eq:u}), we have $F^{(k)} \cup G^{(k)} = C_n^{(k)} =
( C_n^{(k)} \cap S_n^{(k)}) \cup ( C_n^{(k)} \setminus S_n^{(k)})$.
Therefore, (\ref{eq:csg}) and (\ref{eq:csf}) imply that
$F^{(k)} = C_n^{(k)} \cap S_n^{(k)}$, and so
\begin{equation}
F^{(k)} \subseteq S_n^{(k)} . 
\label{eq:fs}
\end{equation}

To continue our argument, we will next show that 
\begin{equation}
F^{(m)} = C_n \; \mbox{and} \; G^{(m)} = \emptyset .  \label{eq:fcg}
\end{equation}
To arrive at a contradiction, assume
$v \in G^{(m)}$. Then there is a string $s \in S_n^{(k)}$ which is not
the prefix of any codeword of $C_n$.  By Theorem~\ref{thm:rightarrow}, 
$v$ has a prefix in $C_n^{(k)}$, say $c$.  Since $v \in G^{(m)}$, it follows
that $c \in C_n^{(k)} \setminus S_n^{(k)}$. By (\ref{eq:cs}) we have
$|v| \geq |c| \geq |s|$.  There are two cases to consider:
\begin{enumerate}
\item $|v| > |s|$:  Since $s$ is a palindrome which is not the prefix
of any codeword in $C_n$, we have that $(C_n \setminus \{v\}) \cup \{s\}$ is a
symmetric fix-free code with $n$ codewords which is better than $C_n$ for any 
probabilistic source.  Hence, $C_n \not\in \mathbb{O}_{n}$, which contradicts 
our assumption. 
\item $|v| = |s|$:  Then $v=c$ and so $v \in C_n^{(k-1)} 
\cup {\cal N}_n (w_{k-1}) \setminus \{w_{k-1} \}$ and $v \not\in S_n^{(k)}$.
Therefore $(S_n^{(k)} \setminus \{s\}) \cup \{v\}$ has the same length
sequence as $S_n^{(k)}$ and greater overlap with $C_n$, which contradicts
our assumption about the choice of $S_n^{(k)}$.
\end{enumerate}

We next show that $w_i \in F^{(i)}$ for some 
$i \in \{k, \ k+1, \ \dots , m-1 \}$.  Suppose that $w_i \not\in F^{(i)}$ for 
all $i \geq k$.  Then by (\ref{eq:ff}) and (\ref{eq:fs}), 
\begin{equation}
F^{(m)} \subseteq \dots \subseteq F^{(k)} \subseteq S_n^{(k)} . 
\label{eq:ffs}
\end{equation}
By (\ref{eq:fcg}), (\ref{eq:ffs}), and the fact that $C_n, \  S_n^{(k)} \in 
\mathbb{S}_{n}$, we obtain $C_n = S_n^{(k)}$, which contradicts our
assumption.

Define the set $\{ i_k , \ \dots , \ i_{k+d-1} \} \subseteq 
\{k, \ \dots , m-1 \}$ to be the collection of indices for which 
$w_{i_l} \in F^{(i_l)}, \; l \in \{k, \ \dots , \ k+d-1\}$ and 
$w_i \in G^{(i)}, \; i \not\in \{i_k, \ \dots , \ i_{k+d-1} \}$. 
Then by (\ref{eq:ff}) and (\ref{eq:fs}), we obtain
\begin{eqnarray}
F^{(i_k)} \subseteq & \dots & \subseteq F^{(k)} \subseteq S_n^{(k)} 
\label{eq:first} \\
F^{(i_{l+1})} \subseteq & \dots & \subseteq F^{(i_l + 1)} , \;
l \in \{k, \ \dots , \ k+d-2 \} \nonumber \\
F^{(m)} \subseteq & \dots & \subseteq F^{(i_{k+d-1} + 1)} \label{eq:last} 
\end{eqnarray}
Since $C_n^{(i_l + 1)} = F^{(i_l + 1)} \cup G^{(i_l + 1)} 
\subseteq ((F^{(i_l)} \setminus \{ w_{i_l} \} ) \cup {\cal N}_n (w_{i_l}))
\cup G^{(i_l)} $ and $w_{i_l} \in F^{(i_l)}$ implies that every element of
${\cal N}_n (w_{i_l})$ has a prefix in $S_n^{(k)}$, we find that
\begin{equation}
F^{(i_l + 1)} 
\subseteq (F^{(i_l)} \setminus \{ w_{i_l} \} ) \cup {\cal N}_n (w_{i_l}),
\; l \in \{k, \ \dots , \ k+d-2 \} \label{eq:fil} .
\end{equation}
From (\ref{eq:first}), we obtain $w_{i_k} \in F^{(i_k)} \subseteq S_n^{(k)} .$
By (\ref{eq:first}) and (\ref{eq:fil}), we can verify that
\begin{displaymath}
F^{(i_k + 1)} 
\subseteq (S_n^{(k)} \setminus \{ w_{i_k} \} ) \cup {\cal N}_n (w_{i_k}).
\end{displaymath}
Therefore, there exists a symmetric fix-free code 
$\hat{S}_n^{(k+1)} \in \mathbb{S}_{n}$ such that
 \begin{equation}
F^{(i_k + 1)} \subseteq \hat{S}_n^{(k+1)} 
\subseteq (S_n^{(k)} \setminus \{ w_{i_k} \} ) \cup {\cal N}_n (w_{i_k}),
\label{eq:k}
\end{equation}
and so $S_n^{(k)} \rightarrow \hat{S}_n^{(k+1)}$.  
Similarly, we can construct a sequence of symmetric fix-free codes  \\
$\hat{S}_n^{(k+2)} , \ \dots , \hat{S}_n^{(k+d)} \in \mathbb{S}_{n}$ for which
 \begin{equation}
F^{(i_l + 1)} \subseteq \hat{S}_n^{(l+1)} 
\subseteq (\hat{S}_n^{(l)} \setminus \{ w_{i_l} \} ) 
\cup {\cal N}_n (w_{i_l}), \; l \in \{ k+1, \ \dots , \ k+d-1 \}.
\label{eq:l}
\end{equation}
Hence, $S_n^{(k)} \rightarrow \hat{S}_n^{(k+1)} \rightarrow \dots \rightarrow
\hat{S}_n^{(k+d)} .$

By (\ref{eq:fcg}), (\ref{eq:last}), and (\ref{eq:l}), we can show that
$C_n = F^{(m)} \subseteq \hat{S}_n^{(k+d)}$.
Because $C_n , \ \hat{S}_n^{(k+d)} \in \mathbb{S}_{n}$,
we have $C_n = \hat{S}_n^{(k+d)}$.  Thus,
\begin{displaymath}
S_n^{(k)} \rightarrow {\hat{S}_n}^{(k+1)} \rightarrow {\hat{S}_n}^{(k+2)} 
\rightarrow \dots \rightarrow {\hat{S}_n}^{(k+d)} = C_n
\end{displaymath}
with $k+d \leq m$.
\end{IEEEproof}
To reiterate the result, if $k-1 \neq m$ we can alter the generation
of code $C_n$ from
\begin{eqnarray*}
R_n = C_n^{(0)} \Rightarrow \dots \Rightarrow C_n^{(k-1)} & \rightarrow & 
C_n^{(k)} \rightarrow \dots \rightarrow C_n^{(m)} = C_n \\ 
\mbox{to } \; R_n = C_n^{(0)} \Rightarrow \dots \Rightarrow C_n^{(k-1)} 
& \Rightarrow & S_n^{(k)} \rightarrow \dots \rightarrow \hat{S}_n^{(k+d)} 
= C_n 
\end{eqnarray*}
for some $k+d \leq m$.  By repeatedly applying this argument we obtain the 
result.
\end{IEEEproof}
{\em Comment:} There is some evidence that for codes in $\mathbb{O}_{n}$ 
the number $m$ of $\Rightarrow$ operations needed is 
$O( n \log_2 n)$.  In \cite[Prop.~2.6]{stak} we showed that the average number
of bits per symbol of the optimal symmetric fix-free code is at most
$2 {\cal H}+1$, where ${\cal H}$ is the binary entropy of the source.
Suppose the source probabilities are $p_1 \geq p_2 \geq \dots \geq p_n$.
Then $m \leq \sum_{i=1}^n (l_i -1) \leq \sum_{i=1}^n p_i (l_i -1) / p_n 
\leq 2{\cal H} / p_n.$
\vspace*{-.1in}
\section{Simplifying the Search for Optimal Symmetric Fix-Free Codes}  
\vspace*{-.1in}
The sequence of symmetric fix-free codes from the root code $R_n$ to an
optimal code $C_n \in \mathbb{O}_{n}$ as defined in 
Theorem~\ref{thm:Rightarrow} is often not unique.  The following result
further specifies such codes.
\begin{lemma}
For any code $C_n \in \mathbb{O}_{n}$, suppose
$R_n = S_n^{(0)} \overset{\pi_1}\Rightarrow S_n^{(1)} 
\overset{\pi_2}\Rightarrow S_n^{(2)} \overset{\pi_3}\Rightarrow 
\dots \overset{\pi_m}\Rightarrow S_n^{(m)} = C_n .$
Then this is a shortest sequence of symmetric fix-free codes transforming
$R_n$ to $C_n$ via repeated uses of the $\Rightarrow$ operation if and only
if $\pi_i$ is a prefix of at least one codeword in $C_n$ for each 
$i \in \{1, \ \dots , \ m\}$.
\label{lem:shortest}
\end{lemma}
\begin{IEEEproof}
Let us first consider the case where the condition is not satisfied.
Let $l \in \{1, \ \dots , \ m\}$ denote the maximum index for which
$\pi_l$ is not a prefix of any codeword in $C_n$.  Observe that it is
impossible to have $l=m$ because $C_n = S_n^{(m)}$ has a nonempty 
intersection with ${\cal N}_n (\pi_m)$ since $C_n$ and $S_n^{(m-1)}$ 
both have $n$ codewords.  Therefore, $l<m$.  For 
$i \geq l+1, \; \pi_i$ is a prefix of at least one
codeword in $C_n$, so $\pi_l$ cannot be a prefix of $\pi_i$.
Thus, by the definition of the $\Rightarrow$ operation we can write
\begin{equation}
S_n^{(i)} \setminus {\cal N}_n (\pi_l)
\subseteq (S_n^{(i-1)} \setminus {\cal N}_n (\pi_l)) \cup {\cal N}_n (\pi_i)
\setminus \{\pi_i \}, \; i \in \{ l+1, \ \dots , \ m\} . \label{eq:a}
\end{equation}
Since $\pi_l$ is not a prefix of $\pi_i, \; i \in \{ l+1, \ \dots , \ m\},$
it follows from (\ref{eq:a}) that for $i \geq l+1$,
\begin{equation}
\pi_i \in S_n^{(i-1)} \setminus {\cal N}_n (\pi_l). \label{eq:B1}
\end{equation}

We will use induction to establish the existence of 
codes \mbox{${C}_n^{(l+1)}, \ \dots , \ 
{C}_n^{(m)} = {{C}_n} \in \mathbb{S}_{n}$} satisfying
\begin{eqnarray}
& & S_n^{(i)} \setminus {\cal N}_n (\pi_l) \subseteq 
C_n^{(i)} , \; i \in \{ l+1, \ \dots , \ m\} , \label{eq:C1} \\
\mbox{and} & & S_n^{(l-1)} \overset{\pi_{l+1}}\rightarrow C_n^{(l+1)} 
\overset{\pi_{l+2}}\rightarrow C_n^{(l+2)} 
\dots \overset{\pi_m}\rightarrow C_n^{(m)} = C_n . \label{eq:d}
\end{eqnarray}
For the basis step, the definition of the $\Rightarrow$ operation implies
\begin{equation}
S_n^{(l)} \setminus {\cal N}_n (\pi_l) \subseteq S_n^{(l-1)} \setminus 
\{\pi_l \}. \label{eq:e}
\end{equation}
Furthermore, we have seen that $\pi_l$ is not a prefix of $\pi_{l+1}$.
By (\ref{eq:B1}) and (\ref{eq:e}) we have
\begin{equation}
\pi_{l+1} \in S_n^{(l-1)} . \label{eq:F1}
\end{equation}
It follows from (\ref{eq:a}) that
\begin{equation}
S_n^{(l+1)} \setminus {\cal N}_n (\pi_l)
\subseteq (S_n^{(l)} \setminus {\cal N}_n (\pi_l)) \cup {\cal N}_n (\pi_{l+1})
\setminus \{\pi_{l+1} \}
\subseteq S_n^{(l)} \cup {\cal N}_n (\pi_{l+1}) \setminus \{\pi_{l+1} \} .
\label{eq:G1}
\end{equation}
Observe that $S_n^{(l+1)} \setminus {\cal N}_n (\pi_l)$
contains at most $n$ words and
$S_n^{(l)} \cup {\cal N}_n (\pi_{l+1}) \setminus \{\pi_{l+1} \}$ 
contains at least $n$ words.  Therefore, by (\ref{eq:F1}) and (\ref{eq:G1}),
there exists ${C}_n^{(l+1)} \in \mathbb{S}_{n}$ such that
$S_n^{(l+1)} \setminus {\cal N}_n (\pi_l) \subseteq C_n^{(l+1)}$
and $S_n^{(l-1)} \overset{\pi_{l+1}}\rightarrow C_n^{(l+1)}$. 

For the inductive step, suppose that for some $l+1 \leq k < m$ we have found
symmetric fix-free codes ${C}_n^{(l+1)}, \ \dots , \ 
{C}_n^{(k)} \in \mathbb{S}_{n}$ which satisfy (\ref{eq:C1}) and
$S_n^{(l-1)} \overset{\pi_{l+1}}\rightarrow C_n^{(l+1)} 
\overset{\pi_{l+2}}\rightarrow C_n^{(l+2)} 
\dots \overset{\pi_k}\rightarrow C_n^{(k)}$.
We next generate $C_n^{(k+1)}$.  By (\ref{eq:B1}), (\ref{eq:C1}), and 
(\ref{eq:a}) we have
\begin{eqnarray*}
& & \pi_{k+1} \in S_n^{(k)} \setminus {\cal N}_n (\pi_l) \subseteq C_n^{(k)}
\; \mbox{and}  \\
& & S_n^{(k+1)} \setminus {\cal N}_n (\pi_l)
\subseteq (S_n^{(k)} \setminus {\cal N}_n (\pi_l)) \cup {\cal N}_n (\pi_{k+1})
\setminus \{\pi_{k+1} \}
\subseteq C_n^{(k)} \cup {\cal N}_n (\pi_{k+1}) \setminus \{\pi_{k+1} \} .
\end{eqnarray*}
Like the argument for the basis step, there exists 
${C}_n^{(k+1)} \in \mathbb{S}_{n}$ for which
$S_n^{(k+1)} \setminus {\cal N}_n (\pi_l) \subseteq C_n^{(k+1)}$ and
$S_n^{(l-1)} \overset{\pi_{l+1}}\rightarrow C_n^{(l+1)} 
\overset{\pi_{l+2}}\rightarrow C_n^{(l+2)} 
\dots \overset{\pi_{k+1}}\rightarrow C_n^{(k+1)}$.  At $k+1=m$ we have
$C_n = S_n^{(m)} \setminus {\cal N}_n (\pi_l)$,
and therefore $C_n = S_n^{(m)} = C_n^{(m)}$.

We have established a sequence of symmetric fix-free codes
$S_n^{(0)}, \ S_n^{(1)}, \ \dots S_n^{(l-1)},$ $C_n^{(l+1)}, \ \dots ,
\ C_n^{(m)}$ for which 
$R_n = S_n^{(0)} \overset{\pi_1}\Rightarrow S_n^{(1)} 
\overset{\pi_2}\Rightarrow S_n^{(2)} \overset{\pi_3}\Rightarrow 
\dots \overset{\pi_{l-1}}\Rightarrow S_n^{(l-1)} 
\overset{\pi_{l+1}}\Rightarrow C_n^{(l+1)} \overset{\pi_{l+2}}\rightarrow 
\dots \overset{\pi_m}\rightarrow C_n^{(m)} = C_n .$
By the argument used in the proof of Theorem~\ref{thm:Rightarrow}, these 
relations imply the existence a sequence of symmetric fix-free codes
$S_n^{(0)}, \ D_n^{(1)} , \ D_n^{(2)}, \ \dots D_n^{(j)} = C_n 
\in \mathbb{S}_{n}$ with $j \leq m-1$ for which
$R_n = S_n^{(0)} \Rightarrow D_n^{(1)} \Rightarrow D_n^{(2)} \Rightarrow 
\dots \Rightarrow D_n^{(j)} = C_n ,$ which demonstrates that
$R_n = S_n^{(0)}, \ S_n^{(1)} , \ S_n^{(2)}, \ \dots S_n^{(m)} = C_n$ 
is not a shortest sequence of codes transforming $R_n$ to $C_n$ via repeated
uses of the $\Rightarrow$ operation. 

For the converse, given an arbitrary code $C_n \in \mathbb{S}_{n}$ 
let $C^{\mbox{{\small prefix}}}$ be the set of palindromes (not including 1) 
which are proper prefixes of at least one codeword in $C_n$.
Suppose we are given a set of codes 
$S_n^{(0)}, \ S_n^{(1)} , \ S_n^{(2)}, \ \dots S_n^{(m)} \in \mathbb{S}_{n}$ 
and palindromes $\{ \pi_1, \ \dots , \pi_m \}$ defined by
$R_n = S_n^{(0)} \overset{\pi_1}\Rightarrow S_n^{(1)} 
\overset{\pi_2}\Rightarrow S_n^{(2)} \overset{\pi_3}\Rightarrow 
\dots \overset{\pi_m}\Rightarrow S_n^{(m)} = C_n .$  We will show that 
$C^{\mbox{{\small prefix}}} \subseteq \{ \pi_1, \ \dots , \pi_m \}$.
 
For each $w \in C_n$, define $C^{\mbox{{\small prefix}}} (w)$ to be the
set of palindromes (not including 1) which are proper prefixes of $w$.
Then $C^{\mbox{{\small prefix}}} = \cup_{w \in C_n}  
C^{\mbox{{\small prefix}}} (w)$.  If $w \in R_n$, then
$C^{\mbox{{\small prefix}}} (w) = \emptyset$. 
Otherwise, there is an ordering of the $\eta_w \geq 1$ strings in
$C^{\mbox{{\small prefix}}} (w)$, say $\sigma_w^{(1)}, \ \dots , \
\sigma_w^{(\eta_w)}$, so that $\sigma_w^{(1)} \in R_n , \
\sigma_w^{(i+1)} \in {\cal N}_n ( \sigma_w^{(i)} )$
for $i \in \{1, \ \dots , \ \eta_w - 1 \}$, and 
$w \in {\cal N}_n ( \sigma_w^{(\eta_w)} )$.  
Observe that $w \in C_n$ implies that $\sigma_w^{(i)} \in
\{ \pi_1, \ \dots , \pi_m \}$ for all $w \not\in R_n$ and
$i \in \{1, \ \dots , \ \eta_w \}$.  Therefore
$C^{\mbox{{\small prefix}}} (w) \subseteq \{ \pi_1, \ \dots , \pi_m \}$ 
for all $w \in C_n$, and so 
\begin{equation}
C^{\mbox{{\small prefix}}} \subseteq \{ \pi_1, \ \dots , \pi_m \}. \label{eq:H1}
\end{equation}

Because $C^{\mbox{{\small prefix}}}$ is determined only by $C_n$, in order for
$S_n^{(0)}, \ S_n^{(1)} , \ S_n^{(2)}, \ \dots S_n^{(m)} \in \mathbb{S}_{n}$ 
to be a shortest sequence of codes transforming
$R_n$ to $C_n$ via uses of the  $\Rightarrow$ operation, it suffices to show
that
\begin{equation}
C^{\mbox{{\small prefix}}} = \{ \pi_1, \ \dots , \pi_m \}. \label{eq:I1}
\end{equation}
The assumption $\{ \pi_1, \ \dots , \pi_m \} \subseteq 
C^{\mbox{{\small prefix}}}$ together with (\ref{eq:H1})
results in (\ref{eq:I1}).
\end{IEEEproof}

Given Lemma~\ref{lem:shortest} and (\ref{eq:I1}), we next show
\begin{theorem}
For any code $C_n \in \mathbb{O}_{n}$, suppose
$R_n = S_n^{(0)} \overset{\pi_1}\Rightarrow S_n^{(1)} 
\overset{\pi_2}\Rightarrow S_n^{(2)} \overset{\pi_3}\Rightarrow 
\dots \overset{\pi_m}\Rightarrow S_n^{(m)} = C_n $ is
a shortest sequence of codes in $\mathbb{S}_{n}$ transforming
$R_n$ to $C_n$ via uses of the  $\Rightarrow$ operation.
Define $C^{\mbox{{\small prefix}}} = \{ \pi_1, \ \dots , \pi_m \}.$
Then any ordering $\sigma_1 , \ \sigma_2, \ \dots , \ \sigma_m$ of the
elements of $C^{\mbox{{\small prefix}}}$ with $i<j$ whenever
$\sigma_i$ is a prefix of $\sigma_j$ corresponds to a sequence of
symmetric fix-free codes 
$C_n^{(\Sigma , 0)} , \ C_n^{(\Sigma , 1)} , \ C_n^{(\Sigma , 2)}
, \ \dots , \ C_n^{(\Sigma , m)} \in \mathbb{S}_{n}$ satisfying
$R_n = C_n^{(\Sigma , 0)} \overset{\sigma_1}\Rightarrow C_n^{(\Sigma , 1)} 
\overset{\sigma_2}\Rightarrow C_n^{(\Sigma , 2)} \overset{\sigma_3}\Rightarrow 
\dots \overset{\sigma_m}\Rightarrow C_n^{(\Sigma , m)} = C_n $. 
\label{thm:Sigma}
\end{theorem}
\begin{IEEEproof}
There are two main parts to the proof.  In the first we show that
there is a set of transformations starting from $\{ \pi_1, \ \dots , \pi_m \}$
and ending in $\{ \sigma_1, \ \dots , \sigma_m \}$ which at each step
involves a transposition of an adjacent pair of strings while maintaining
the invariant that any palindrome (not including 1) which is a proper prefix 
of a palindrome in the list always precedes it.  In the second part we consider
the effect of a (valid) transposition of an adjacent pair of strings in
devising shortest transformation from $R_n$ to $C_n \in \mathbb{O}_{n}$ 
via uses of the $\Rightarrow$ operation.

For the first part of the proof, for a sequence (of numbers or strings) \\
$A= (a_1, \ a_2, \ \dots , \ a_m)$, define $A_i, \ 
i \in \{1, \ \dots , \ m-1\}$,
as the permutation of $A$ obtained by transposing $a_i$ and $a_{i+1}$.
For example, if $A=(1,2,3,4)$, then  $A_1=(2,1,3,4), \ A_2=(1,3,2,4), \
A_3=(1,2,4,3)$.  We have the following result.
\begin{lemma}
For $( \pi_1, \ \dots , \pi_m )$ and $( \sigma_1, \ \dots , \sigma_m )$ defined
in Theorem~\ref{thm:Sigma}, define $\Omega^0 = ( \pi_1, \ \dots , \pi_m )$.
Then there is a number $k < m^2$, a sequence of indices
$a_1 , \ \dots , \ a_k \in \{1, \ \dots , \ m-1\}$, and 
a sequence of pairwise permutations
starting from $\Omega^0$ with $\Omega^i = (\Omega^{i-1})_{a_i}$ and  
$\Omega^k = ( \sigma_1, \ \dots , \sigma_m )$ 
such that for all $i, \ \Omega^i$ satisfies the constraint that the proper
prefixes in the list of each palindrome precede it in the ordering.
\label{lem:permutations}
\end{lemma}
\begin{IEEEproof}
Suppose we know $\Omega^0, \ \dots , \ \Omega^i = (w_1^i, \ \dots , \ w_m^i)$,
and we wish to construct $\Omega^{i+1}$.  Let $h_i$ be the maximum index for 
which $w_g^i \neq \sigma_g$.  Then there is some $l_i < h_i$ for which 
$w_{l_i}^i = \sigma_{h_i}$.  We claim that we can choose
$\Omega^{i+1} = (\Omega^i)_{l_i}$; i.e., $w_{l_i}^i$ is not a prefix
of $w_{l_i+1}^i$.  This is clearly true if $\sigma_{h_i}$ is not a prefix
of $\sigma_j, \ j \neq h_i$.  If $\sigma_{h_i}=w_{l_i}^i$ 
is a proper prefix of some $\sigma_j=w_{l_i+1}^i$, then by assumption 
$j > h_i$, and hence $h_i$ is not the maximum index for which
$w_g^i \neq \sigma_g$.  

Given this choice of $\Omega^{i+1}$, let us consider the ordered pair
$(l_{i+1}, \ h_{i+1})$.  If $l_i+1 < h_i$, then 
$(l_{i+1}, \ h_{i+1}) = (l_i+1, \ h_i)$, and if $l_i+1 = h_i$, then 
$h_{i+1} < h_i$.  Since $(l_i, \ h_i) \neq (l_j, \ h_j)$ for $i \neq j$,
eventually the sequence of pairwise permutations will terminate in
$\Omega^k = ( \sigma_1, \ \dots , \sigma_m )$.  
\end{IEEEproof}

For the second part of the proof of Theorem~\ref{thm:Sigma}, we are given
that for $\Omega^0, R_n = S_n^{(0)} \overset{\pi_1}\Rightarrow S_n^{(1)} 
\overset{\pi_2}\Rightarrow S_n^{(2)} \overset{\pi_3}\Rightarrow 
\dots \overset{\pi_m}\Rightarrow S_n^{(m)} = C_n $ is
a shortest sequence of codes in $\mathbb{S}_{n}$ transforming
$R_n$ to $C_n$ via uses of the  $\Rightarrow$ operation.  Next suppose that 
for some $i \geq 0$, there is a sequence of symmetric fix-free codes
$C_n^{(\Omega^i , 0)} , \ C_n^{(\Omega^i , 1)} , \ C_n^{(\Omega^i , 2)}
, \ \dots , \ C_n^{(\Omega^i , m)} \in \mathbb{S}_{n}$ satisfying
$R_n = C_n^{(\Omega^i , 0)} \overset{w_1^i}\Rightarrow C_n^{(\Omega^i , 1)} 
\overset{w_2^i}\Rightarrow C_n^{(\Omega^i , 2)} \overset{w_3^i}\Rightarrow 
\dots \overset{w_m^i}\Rightarrow C_n^{(\Omega^i , m)} = C_n $. 
By Lemma~\ref{lem:permutations}, to complete the proof of 
Theorem~\ref{thm:Sigma} it suffices to show that
there is a sequence of symmetric fix-free codes
$C_n^{(\Omega^{i+1} , 0)} , \ C_n^{(\Omega^{i+1} , 1)} , \ 
C_n^{(\Omega^{i+1} , 2)} , \ \dots , \ C_n^{(\Omega^{i+1} , m)} 
\in \mathbb{S}_{n}$ satisfying $R_n = C_n^{(\Omega^{i+1} , 0)} 
\overset{w_1^{i+1}}\Rightarrow C_n^{(\Omega^{i+1} , 1)} 
\overset{w_2^{i+1}}\Rightarrow C_n^{(\Omega^{i+1} , 2)} 
\overset{w_3^{i+1}}\Rightarrow \dots \overset{w_m^{i+1}}\Rightarrow 
C_n^{(\Omega^{i+1} , m)} = C_n $. 

From the proof of Lemma~\ref{lem:permutations}, we have the following
relationship between $\Omega^{i+1} = (w_1^{i+1}, \ \dots , \ w_m^{i+1})$
and $\Omega^i = (w_1^i, \ \dots , \ w_m^i)$:
\begin{displaymath}
w_j^{i+1} = \left\{ \begin{array}{ll}
w_j^i , & j \not\in \{l_i, l_{i+1} \} \\
w_{l_i+1}^i , & j=l_i \\
w_{l_i}^i , & j=l_i+1 \end{array} \right.
\end{displaymath}
In the proof of Lemma~\ref{lem:permutations} we argued that $w_{l_i}^i$
is not a prefix of $w_{l_i+1}^i$ (or vice versa).  Therefore, for $j<l_i$
we will choose $C_n^{(\Omega^{i+1} , j)} = C_n^{(\Omega^i , j)}$.
If there exists $C_n^{(\Omega^{i+1} , l_i)} \in \mathbb{S}_{n}$ 
for which 
\begin{equation}
C_n^{(\Omega^i , l_i-1)} 
\overset{w_{l_i+1}^i}\Rightarrow C_n^{(\Omega^{i+1} , l_i)} 
\overset{w_{l_i}^i}\Rightarrow C_n^{(\Omega^i , l_i+1)} ,
\label{eq:J}
\end{equation}
then for $j \geq l_i +1$ we can choose $C_n^{(\Omega^{i+1} , j)} = 
C_n^{(\Omega^i , j)}$.  We next establish the existence of
$C_n^{(\Omega^{i+1} , l_i)}$ to satisfy (\ref{eq:J}).
To simplify notation, define
\begin{displaymath}
S_n = C_n^{(\Omega^i , l_i-1)} , 
\ I_n = C_n^{(\Omega^i , l_i)} , \ S_n^{'} = C_n^{(\Omega^i , l_i+1)} , \
\omega_1 = w_{l_i} , \ \omega_2 = w_{l_i+1} 
\end{displaymath}
so that
\begin{equation}
S_n \overset{\omega_1}\Rightarrow I_n \overset{\omega_2}\Rightarrow S_n^{'} .
\label{eq:K}
\end{equation}
Let $C_n (\omega_1)$ be the subset of words in $C_n$ which have $\omega_1$
as a prefix.  By Lemma~\ref{lem:shortest}, $C_n (\omega_1) \neq \emptyset .$ 
Since $C_n$ and $S_n$ both have $n$ strings, there exists 
$S_n (\omega_1) \subseteq S_n$ with 
$|S_n (\omega_1)|=|C_n (\omega_1)|, \ \omega_1 \in S_n (\omega_1)$,
and $\omega \in S_n (\omega_1)$ is not a prefix of any codeword in $C_n$
if $\omega \neq \omega_1$.
Observe that 
$(C_n \setminus C_n (\omega_1)) \cup S_n (\omega_1) \in \mathbb{S}_{n}$. 
To arrive at a contradiction, suppose 
$\min_{\sigma\in\mathcal{N}_{n}\left( \omega_1 \right)}\left|\sigma\right|\geq\max_{\sigma\in S_n}\left|\sigma\right|.$ Then
$\min_{\sigma\in C_{n}\left( \omega_1 \right)}\left|\sigma\right|\geq\max_{\sigma\in S_n (\omega_1)}\left|\sigma\right|$
and $\min_{\sigma\in C_{n}\left( \omega_1 \right)}\left|\sigma\right|
\geq |\omega_1| + 1$.
Therefore the code 
$(C_n \setminus C_n (\omega_1)) \cup S_n (\omega_1)$ is a strictly better 
symmetric fix-free code than $C_n$ for any choice of source probabilities, 
contradicting the assumption that $C_n \in \mathbb{O}_{n}$.
Hence,
\begin{equation}
\min_{\sigma\in\mathcal{N}_{n}\left(\omega_1 \right)}\left|\sigma\right|<\max_{\sigma\in S_n}\left|\sigma\right|. \label{eq:L1}
\end{equation}
We likewise have
\begin{equation}
\min_{\sigma\in\mathcal{N}_{n}\left(\omega_2 \right)}\left|\sigma\right|<\max_{\sigma\in I_n}\left|\sigma\right|. \label{eq:L2}
\end{equation}
Since $\omega_1 \in S_n$ is a prefix of at least one codeword in $C_n$,
it must also be a prefix of at least one codeword in $S_n^{'}$.
Furthermore, because $\omega_1, \omega_2 \in S_n$ and are distinct,
$\omega_1$ is not a prefix of 
any string in $\mathcal{N}_{n}\left(\omega_2 \right)$.
Hence,
\begin{equation}
\mathcal{N}_{n}\left(\omega_1 \right) \cap S_n^{'} \neq \emptyset .
\label{eq:M}
\end{equation}
In order to continue our discussion of the transposition of a successive
pair of $\Rightarrow$ operations, we introduce the following notation:
\begin{eqnarray*}
I_n & = & \hat{I} (\omega_1) \cup \hat{S} (\omega_1) \\
\hat{S} (\omega_1) & \subseteq & S_n \setminus\left\{ \omega_1 \right\} \\
\hat{I} (\omega_1) & \subseteq & \mathcal{N}_{n}\left( \omega_1 \right) \\
S_n^{'} & = & \tilde{S} (\omega_1 , \omega_2) \cup \hat{J} (\omega_1) 
\cup \tilde{J} (\omega_2) \\
\tilde{S} (\omega_1 , \omega_2) & \subseteq & \hat{S} (\omega_1) \setminus
\{ \omega_2 \}  \subseteq S_n \setminus\left\{ \omega_1 , \omega_2 \right\} \\
\hat{J} (\omega_1) & \subseteq & \hat{I} (\omega_1) \subseteq 
\mathcal{N}_{n}\left( \omega_1 \right) \\
\tilde{J} (\omega_2) & \subseteq & \mathcal{N}_{n}\left( \omega_2 \right) 
\end{eqnarray*}

We have the following result.
\begin{prop}
\label{prop:existence}
There exists $J_n \in\mathbb{S}_{n}$ such that
$\tilde{S} (\omega_1 , \omega_2) \cup \{ \omega_1 \} 
\cup \tilde{J} (\omega_2) \subseteq J_n$
and $S_n \overset{\omega_2}{\Rightarrow}J_n .$\end{prop}
\begin{IEEEproof}
By (\ref{eq:K}), $I_n\overset{\omega_2}{\Rightarrow}S_n^{'}$, and it follows 
that $\mathcal{N}_{n}\left( \omega_2 \right) \neq \emptyset$.
Therefore there is at least one choice for $I_n^{'}\in\mathbb{S}_{n}$ 
for which
\begin{equation}
S_n \overset{\omega_2}{\Rightarrow} I_n^{'}.\label{eq:a_0}
\end{equation}
We will next show that $\omega_1 \in I_n^{'}$.  To arrive at a contradiction,
suppose $\omega_1 \not\in I_n^{'}$.  
Then by the definition of the $\Rightarrow$ operation
\begin{equation}
\left| \omega_1 \right| \geq \max_{\sigma\in I_n^{'}} \left|\sigma\right|.
\label{eq:a_1}
\end{equation}
Define sets $S^{\star} (\omega_2)$ and $J^{\star} (\omega_2)$ by
\begin{eqnarray*}
I_n^{'} & = & S^{\star} (\omega_2) \cup J^{\star} (\omega_2) \\
S^{\star} (\omega_2) & \subseteq & S_n \setminus\left\{ \omega_2 \right\} \\
J^{\star} (\omega_2) & \subseteq & \mathcal{N}_{n}\left( \omega_2 \right) 
\end{eqnarray*}
Since $S^{\star} (\omega_2) \subseteq I_n^{'}$, (\ref{eq:a_1}) implies
\begin{equation}
\left| \omega_1 \right|\geq\max_{\sigma\in S^{\star} (\omega_2)}
\left|\sigma\right|.\label{eq:a_2}
\end{equation}
The relation $S_n \overset{\omega_1}{\Rightarrow}I_n$ implies that $I_n$ 
contains all elements of $S_n$ with length at most $\left| \omega_1 \right|$,
and combined with (\ref{eq:a_2}) we obtain
$S^{\star} (\omega_2) \subseteq I_n \setminus\left\{ \omega_2 \right\} .$ 
Thus, 
\[
I_n^{'}=S^{\star} (\omega_2) \cup J^{\star} (\omega_2) 
\subseteq \left(I_n \setminus\left\{ \omega_2 \right\} \right) \cup
\mathcal{N}_{n}\left( \omega_2 \right).
\]
The previous relation and (\ref{eq:K}) imply 
\begin{equation}
I_n \overset{\omega_2}{\rightarrow} I_n^{'} \; \mbox{and} \;
I_n \overset{\omega_2}{\Rightarrow} S_n^{'} .
\label{eq:a_3}
\end{equation}
Thus, the difference between the $\rightarrow$ and $\Rightarrow$
operations, (\ref{eq:a_1}), (\ref{eq:a_3}), and (\ref{eq:M}) imply
\begin{displaymath}
\left| \omega_1 \right| \geq \max_{\sigma\in I_n^{'}} \left|\sigma\right|
\geq \max_{\sigma\in S_n^{'}} \left|\sigma\right|
\geq \min_{\sigma\in {\cal N}_n (\omega_1)} \left|\sigma\right|
> \left| \omega_1 \right| , 
\end{displaymath}
which is impossible.
Hence the assumption that $\omega_1 \not\in I_n^{'}$ 
was false. Therefore
\begin{equation}
S_n \overset{\omega_2}{\Rightarrow} I_n^{'} \; \mbox{implies} \;
\omega_1 \in I_n^{'}. \label{eq:implies}
\end{equation}

Recall that 
$S_n^{'} = \tilde{S} (\omega_1 , \omega_2) \cup \hat{J} (\omega_1) 
\cup \tilde{J} (\omega_2)$.
By (\ref{eq:M}) we have $\hat{J} (\omega_1) \neq \emptyset$.
Since $S_n^{'}$ has $n$ codewords, it follows that
$\tilde{S} (\omega_1 , \omega_2) \cup \{ \omega_1 \} \cup \tilde{J} (\omega_2)$
has at most $n$ elements. To arrive at a contradiction, suppose there is no 
$J_n$ that simultaneously satisfies $S_n \overset{\omega_2}{\Rightarrow}J_n $
and $\tilde{S} (\omega_1 , \omega_2) \cup \{ \omega_1 \} 
\cup \tilde{J} (\omega_2) \subseteq J_n$.
Then choose some set $J_n$ for which $S_n \overset{\omega_2}{\Rightarrow}J_n $.
Since $\tilde{S} (\omega_1 , \omega_2) \cup \{ \omega_1 \} 
\cup \tilde{J} (\omega_2) \not\subseteq J_n$,
the relation $J_n \subseteq S_n \cup\mathcal{N}_{n}\left( \omega_2 \right)
\setminus\left\{ \omega_2 \right\} $
and the definition of the $\Rightarrow$ operation imply the existence of 
$x\in J_n \setminus (\tilde{S} (\omega_1 , \omega_2) \cup 
\{ \omega_1 \} \cup \tilde{J} (\omega_2))$
and $y\in \tilde{S} (\omega_1 , \omega_2) \cup 
\{ \omega_1 \} \cup \tilde{J} (\omega_2) \setminus J_n$
with $\left|y\right|>\left|x\right|.$ 
By (\ref{eq:implies}) we know $\omega_1 \in J_n$, so $x \neq \omega_1$ 
and $y\neq \omega_1.$ 
Therefore, $y\in \tilde{S} (\omega_1 , \omega_2) \cup \tilde{J} (\omega_2)$;
i.e., $y\in S_n^{'}$. Similarly,
$x\in J_n \subseteq S_n \cup\mathcal{N}_{n}\left(\omega_2 \right)\setminus\left\{ \omega_2 \right\} $
and $x\notin (\tilde{S} (\omega_1 , \omega_2) \cup 
\{ \omega_1 \} \cup \tilde{J} (\omega_2))$ implies that
$x\notin S_n^{'}.$ Since $x \in J_n$ and $x\neq \omega_1$ we consider two
exhaustive cases for the membership of $x$:
\begin{itemize}
\item $x\in \hat{S} (\omega_1 ) \cup\mathcal{N}_{n}\left( \omega_2 \right)
\setminus\left\{ \omega_2 \right\} :$
Since $\hat{S} (\omega_1 ) \subseteq I_n$ we have
$x\in I_n \cup\mathcal{N}_{n}\left( \omega_2 \right) \setminus
\left\{ \omega_2 \right\} .$
Thus, there exists $S_n^{''} \in \mathbb{S}_{n}$ such that $x\in S_n^{''}$
and $I_n \overset{\omega_2}{\rightarrow} S_n^{''}.$ 
Recall that $I_n \overset{\omega_2}{\Rightarrow} S_n^{'}.$ 
We saw earlier that $x\notin S_n^{'}$ and $y \in S_n^{'} .$ Therefore,
$\left|x\right|\geq\max_{\sigma\in S_n^{'}}\left|\sigma\right|
\geq\left|y\right|$, which violates our earlier argument that 
$\left|y\right|>\left|x\right|.$ 
\item $x\in S_n \setminus \{ \hat{S} (\omega_1) \cup \left\{ \omega_1 \right\}
\} :$ Since $x\in S_n \setminus \{ \omega_1 \} 
\subseteq S_n \cup \mathcal{N}_{n} \left( \omega_1 \right)
\setminus\left\{ \omega_1 \right\} $,
there exists $I_n^{''}\in\mathbb{S}_{n}$ 
such that $x\in I_n^{''}$ and $S_n \overset{\omega_1}{\rightarrow} I_n^{''}.$
Since $S_n \cap \mathcal{N}_{n}\left(\omega_1 \right) = \emptyset$,
we have $x \not\in  \mathcal{N}_{n}\left(\omega_1 \right)$.
We also assume $x \not\in  \hat{S} \left(\omega_1 \right)$.
It follows that $x\notin I_n$.
Recall that $S_n \overset{\omega_1}{\Rightarrow} I_n .$
Therefore, $\left|x\right|\geq\max_{\sigma\in I_n} |\sigma| .$ 
By (\ref{eq:L2}), $\max_{\sigma\in I_n}\left|\sigma\right| > 
\min_{\sigma\in\mathcal{N}_{n}\left(\omega_2 \right)}\left|\sigma\right|.$
$S_n^{'}$ consists of the smallest elements of
$I_n \cup \mathcal{N}_{n} \left( \omega_2 \right)
\setminus\left\{ \omega_2 \right\} $,
so $\max_{\sigma\in I_n} |\sigma| \geq\max_{\sigma\in S_n^{'}} |\sigma|.$
We have already seen that $y \in S_n^{'}$.  Combining these observations we
obtain 
$\left|x\right|\geq\max_{\sigma\in I_n} \left|\sigma\right|
\geq\max_{\sigma\in S_n^{'}} \left|\sigma\right| \geq\left|y\right|,$
which violates our earlier argument that $\left|y\right|>\left|x\right|.$ 
\end{itemize}
Therefore, our assumption was false, and this establishes the proposition.
\end{IEEEproof}
\begin{prop}
\label{prop:final}For the symmetric fix-free code $J_n$ described by 
Proposition \ref{prop:existence},
\[
J_n \overset{\omega_1 }{\Rightarrow} S_n^{'}.
\]
\end{prop}
\begin{IEEEproof}
\label{final}
Recall that $S_n^{'} = \tilde{S} (\omega_1 , \omega_2) \cup \hat{J} (\omega_1) 
\cup \tilde{J} (\omega_2)$
and $\tilde{S} (\omega_1 , \omega_2) \cup \{ \omega_1 \} 
\cup \tilde{J} (\omega_2) \subseteq J_n$.
Thus, $S_n^{'} \subseteq J_n \cup \mathcal{N}_{n}\left(\omega_1 \right)
\setminus\left\{ \omega_1 \right\} .$
Therefore, $J_n\overset{\omega_1}{\rightarrow} S_n^{'}.$ 
To arrive at a contradiction, suppose 
$J_n \not\Rightarrow S_n^{'}$.  Then choose some
$S_n^{''}$  to satisfy $J_n\overset{\omega_1}{\Rightarrow}S_n^{''}$. 
There exists $x\in S_n^{''}\setminus S_n^{'}$ and 
$y\in S_n^{'} \setminus S_n^{''}$ such that $\left|x\right|<\left|y\right|.$
Observe that $x\in J_n\cup\mathcal{N}_{n}\left( \omega_1 \right) \setminus
\left\{ \omega_1 \right\} \subseteq S_n \cup
\mathcal{N}_{n}\left( \omega_1 \right)\cup\mathcal{N}_{n}\left( \omega_2 
\right)\setminus\left\{ \omega_1 , \omega_2 \right\} .$ 
There are two exhaustive cases for the membership of $x$:
\begin{itemize}
\item $x\in I_n \cup \mathcal{N}_{n}\left(\omega_2 \right)\setminus\left\{ \omega_2 \right\} :$
There exists ${\tilde{S}}_n^{'}\in\mathbb{S}_{n}$ with 
$x\in {\tilde{S}}_n^{'}\setminus S_n^{'}$
and $I_n \overset{\omega_2}{\rightarrow}{\tilde{S}}_n^{'}.$ 
By (\ref{eq:K}), $I_n \overset{\omega_2}{\Rightarrow}{{S}}_n^{'}$.
Since $y \in S_n^{'}$ it follows that
$\left|x\right|\geq\max_{\sigma\in S_n^{'}} |\sigma|
\geq\left|y\right|,$ which contradicts our assumption that $|x|<|y|$.
\item $x\in S_n \cup \mathcal{N}_{n} \left(\omega_1 \right)
\setminus (I_n \cup\left\{ \omega_1 \right\}) :$
Since $x\in S_n \cup \mathcal{N}_{n} \left(\omega_1 \right) \setminus 
\left\{ \omega_1 \right\},$ there exists
$I_n^{''}\in\mathbb{S}_{n}$ such that $x\in I_n^{''}\setminus I_n$
and $S_n \overset{\omega_1}{\rightarrow}I_n^{''}.$ 
By (\ref{eq:K}), $S_n \overset{\omega_1}{\Rightarrow}I_n$.
Since $x \not\in I_n$
we can conclude that $\left|x\right|\geq\max_{\sigma\in I_n} | \sigma |$
and repeat the end of the argument for
Proposition~\ref{prop:existence} to obtain a contradiction.
\end{itemize}
Since our assumption that $J_n \not\Rightarrow S_n^{'}$ was false, we have
established the proposition.
\end{IEEEproof}
To complete the proof of Theorem~\ref{thm:Sigma} we choose
$C_n^{(\Omega^{i+1} , l_i)} = J_n$.
\end{IEEEproof}
{\em Remark:} Lemma~\ref{lem:shortest} and Theorem~\ref{thm:Sigma} 
are important to reduce the computational complexity of the search for 
optimal codes because by allowing a natural ordering to be imposed on 
the strings in $C^{\mbox{{\small prefix}}}$ one can potentially have a large
reduction in the number of sequences of transformations that need to be 
considered.

Thus far we have provided a way to generate any code in $\mathbb{O}_{n}$,
but the procedure will also generate some codes in 
$\mathbb{S}_{n} \setminus \mathbb{O}_{n}$. 
Therefore, it is desirable to provide simple tests to reduce the number of
candidate for codes in $\mathbb{O}_{n}$.
We begin by describing a previously known property of optimal sorted and
nondecreasing sequences of codeword lengths corresponding to symmetric
fix-free codes.  We then offer simplifications of this result,
including a generalization of Theorem~\ref{thm:stak}.

\begin{lemma}
\cite[Lemma~2.1]{stak}
Let $(l_1, \ l_2, \ \dots , \ l_n)$ be the sorted and non-decreasing 
sequence of codeword lengths corresponding to a symmetric fix-free code
and $(l_1^{'}, l_2^{'}, \dots , l_n^{'})$
be a non-decreasing sequence of natural numbers for which 
\begin{displaymath}
\mbox{$\sum_{j=1}^i$} l_j^{'} \; \geq \; 
\mbox{$\sum_{j=1}^i$} l_j \; \mbox{ for each } \;
i \in \{1, \ \dots , \ n\}.
\end{displaymath}
Then $(l_1^{'}, \ l_2^{'}, \ \dots , \ l_n^{'})$ need not be considered as
the potential codeword lengths of an optimal symmetric fix-free code.
\label{lemma:dominant}
\end{lemma}

In the previous result we say length sequence $(l_1, \ l_2, \ \dots , \ l_n)$ 
{\em dominates} the sequence $(l_1^{'}, l_2^{'}, \dots , l_n^{'})$.
Let $\mathbb{D}_{n} \subset \mathbb{S}_{n}$ be the set of symmetric
fix-free codes with sorted and non-decreasing codeword lengths sequences 
each of which is not dominated by the sorted and non-decreasing codeword 
length sequence of any other code in $\mathbb{S}_{n}$. 
We have $\mathbb{O}_{n} \subseteq \mathbb{D}_{n}$,
but it is unknown if $\mathbb{O}_{n} = \mathbb{D}_{n}$ for all $n$.

For symmetric fix-free codes related by the $\Rightarrow$ operation,
the $n$ inequalities of Lemma~\ref{lemma:dominant}
can be reduced to one.  We begin with a special case of this result.

\begin{prop}
Suppose that the code $S_n^{'}$ is a candidate for membership
in $\mathbb{O}_{n}$, and let $S_n \in \mathbb{S}_{n}$ be a code 
in a shortest transformation from $R_n$ to $S_n^{'}$
through a sequence of $\Rightarrow$ operations.
Let $(l_1, \ l_2, \ \dots , \ l_n)$ and 
$(l_1^{'}, \ l_2^{'}, \ \dots , \ l_n^{'})$ be the sorted and non-decreasing 
sequences of codeword lengths of $S_n$ and $S_n^{'}$, respectively.
Suppose that $\sum_{j=1}^n l_j^{'} \; \geq \; \sum_{j=1}^n l_j $. 
If the portion of the shortest transformation from $S_n$
to $S_n^{'}$ satisfies either
\begin{itemize}
\item $S_n \overset{\pi_1}\Rightarrow S_n^{'}$ or 
\item there is a sequence of symmetric fix-free codes 
$S_{n}^{\left(1\right)}, \ S_{n}^{\left(2\right)}, \ \ldots , \
S_{n}^{\left(h \right)}\in\mathbb{S}_{n}$
for some $h \geq 2$ with
\[
S_{n}=S_{n}^{\left(0\right)} \overset{\pi_1}\Rightarrow 
S_{n}^{\left(1\right)} \overset{\pi_2}\Rightarrow S_{n}^{\left(2\right)} 
\overset{\pi_3}\Rightarrow
\cdots \overset{\pi_h}\Rightarrow S_{n}^{\left(h\right)}=S_{n}^{\prime}
\]
and with $\pi_1$ being a prefix of $\pi_i$ for $i \geq 2$,
\end{itemize}
then $S_n^{'} \not\in \mathbb{O}_{n}$. 
\label{prop:sum}
\end{prop}
\begin{IEEEproof}
We begin by considering the first case and later show how to extend the
argument to the second case.

We are given that $S_{n}^{\prime}\subseteq S_{n}\cup\mathcal{N}_{n}
\left(\pi_1 \right)\setminus\left\{ \pi_1 \right\} .$
For integers $\lambda$ let $\tilde{S}_{n}^{\lambda}$ denote the subset of
$S_{n}$ with string lengths greater than $\lambda.$ By the definition of the
$\Rightarrow$ operator, there is some $\lambda$ for which
\begin{equation}
S_{n}^{\prime}\subseteq S_{n}\cup\mathcal{N}_{\lambda}\left(\pi_1 \right)
\setminus \{ \left\{ \pi_1 \right\} \cup \tilde{S}_{n}^{\lambda} \} .
\label{eq:lambda}
\end{equation}
Let 
\begin{eqnarray*}
D & = & S_{n}\setminus S_{n}^{\prime} \\
D^{\prime} & = & S_{n}^{\prime}\setminus S_{n} \\
m & = & \left|D\right|=\left|D^{\prime}\right|.
\end{eqnarray*} 
Let $\left(d_{1},\ldots,d_{m}\right)$
and $\left(d_{1}^{\prime},\ldots,d_{m}^{\prime}\right)$ 
respectively denote the sorted and non-decreasing sequences of codeword 
lengths of $D$ and $D^{\prime}.$  Then
\begin{align}
\left|\pi_1 \right|=d_{1}<\lambda+1\leq d_{2}\leq\ldots\leq d_{m}\label{eq:A}\\
d_{1}+1\leq d_{1}^{\prime}\leq d_{2}^{\prime}\leq\ldots\leq d_{m}^{\prime}\leq\lambda . \label{eq:B}
\end{align}
The condition $\sum_{j=1}^{n}l_{j}^{\prime}\geq\sum_{j=1}^{n}l_{j}$
is equivalent to 
\begin{equation}
d_{1}^{\prime}-d_{1}\geq\left(d_{2}-d_{2}^{\prime}\right)+\ldots+\left(d_{m}-d_{m}^{\prime}\right),\label{eq:C}
\end{equation}
and (\ref{eq:A}) and (\ref{eq:B}) imply that 
\begin{equation}
d_{j}\geq d_{j}^{\prime}+1, \; j\in\left\{ 2, \ \ldots, \ m\right\} . 
\label{eq:D}
\end{equation}
We would like to show that
$\sum_{j=1}^{k}l_{j}^{\prime}\geq\sum_{j=1}^{k}l_{j}, \;
k\in\left\{ 1, \ 2 \ \ldots, \ n\right\}.$
Let $i$ be the largest index for which $l_{i}=d_{1}.$ 
Then the preceding inequality
is an equality for $1\leq k\leq i-1.$ Let $\iota$ be the index for
which $l_{\iota}\leq d_{1}^{\prime}<l_{\iota+1}.$ Then for $i\leq k\leq\iota-1,$
\[
\sum_{j=1}^{k}l_{j}^{\prime}=\sum_{j=1}^{i-1}l_{j}+\sum_{j=i}^{k}l_{j+1}\geq\sum_{j=1}^{k}l_{j}.
\]
For $\iota\leq k\leq n,$ suppose that $l_{1}^{\prime},l_{2}^{\prime},\ldots,l_{k}^{\prime}$
incorporates the $g_{k}$ shortest new codeword lengths 
$d_{1}^{\prime},d_{2}^{\prime},\ldots,d_{g_k}^{\prime}.$
If $g_{k}=1,$ then (\ref{eq:B}) implies 
$\sum_{j=1}^{k}\left(l_{j}^{\prime}-l_{j}\right)=d_{1}^{\prime}-d_{1}\geq1.$
For $2\leq g_{k}\leq m,$ (\ref{eq:C}) and (\ref{eq:D})
imply 
\[
\sum_{j=1}^{k}\left(l_{j}^{\prime}-l_{j}\right)=d_{1}^{\prime}-d_{1}-\sum_{j=2}^{g_{k}}\left(d_{j}-d_{j}^{\prime}\right)\geq0,
\]
as desired.

For the second case, we let $\mathcal{N}_n^{\star} (\sigma)$
denotes the set of all palindromes of length at most $n$ with $\sigma$
as a proper prefix.
The only change needed to the previous discussion is to replace 
(\ref{eq:lambda}) with
\begin{displaymath}
S_{n}^{\prime} \subseteq S_{n}\cup \mathcal{N}_{\lambda^{\star}}^{\star} 
(\pi_1) \setminus \{ \left\{ \pi_1 \right\} \cup 
\tilde{S}_{n}^{\lambda^{\star} } \}
\end{displaymath}
for some $\lambda^{\star}$ and to replace $\lambda$ with $\lambda^{\star}$
in (\ref{eq:A}) and (\ref{eq:B}).  The rest of the proof remains the same
as in the first case.
\end{IEEEproof}

We next extend Proposition~\ref{prop:sum} and simultaneously generalize
Theorem~\ref{thm:stak}.
\begin{theorem}
Consider a code $S_n^{'} \in \mathbb{O}_{n}$, and suppose 
$S_n \in \mathbb{S}_{n}$ is one of the codes in a shortest transformation 
from $R_n$ to $S_n^{'}$ through a sequence of $\Rightarrow$ operations.
Suppose the portion of this shortest transformation from $S_n$ to $S_n^{'}$
involves the sequence of symmetric fix-free codes  
$S_n^{(1)}, \ S_n^{(2)}, \ \dots , \ S_n^{(h)} \in \mathbb{S}_{n}$ 
for some $h \geq 1$ and satisfies
\begin{displaymath}
S_n = S_n^{(0)} \overset{\sigma_1}\Rightarrow S_n^{(1)} 
\overset{\sigma_2}\Rightarrow S_n^{(2)} \overset{\sigma_3}\Rightarrow 
\dots \overset{\sigma_h}\Rightarrow S_n^{(h)} = S_n^{'}.
\end{displaymath}
Let $(l_1, \ l_2, \ \dots , \ l_n)$ and 
$(l_1^{'}, \ l_2^{'}, \ \dots , \ l_n^{'})$ be the sorted and non-decreasing 
sequences of codeword lengths of $S_n$ and $S_n^{'}$, respectively.
Let $\tilde{l}_n^{(i)}, i \in \{ 0, \ 1, \ \dots , \ h \}$, denote the
maximum codeword length of 
$S_n^{(i)}$.  Then $l_n^{'} = \tilde{l}_n^{(h)} \leq \tilde{l}_n^{(h-1)}
\leq \dots \leq \tilde{l}_n^{(1)} \leq \tilde{l}_n^{(0)} = l_{n}$ and 
$\sum_{j=1}^n l_j^{'} \; < \; \sum_{j=1}^n l_j .$
\label{thm:sum}
\end{theorem}
\begin{IEEEproof}
Let $S_n^{'} (\sigma_i)$ be the subset of words in $S_n^{'}$ which have 
$\sigma_i$ as a prefix.  By Lemma~\ref{lem:shortest}, 
$S_n^{'} (\sigma_i) \neq \emptyset .$ 
Since $S_n^{'}$ and $S_n^{(i-1)}$ both have $n$ strings, there exists 
$S_n^{(i-1)}(\sigma_i) \subseteq S_n^{(i-1)}$ with 
$|S_n^{(i-1)} (\sigma_i)|=|S_n^{'} (\sigma_i)|, \ 
\sigma_i \in S_n^{(i-1)} (\sigma_i)$, and $\sigma \in S_n^{(i-1)} (\sigma_i)$ 
is not a prefix of any codeword in $S_n^{'}$ if $\sigma \neq \sigma_i$.
Observe that 
$(S_n^{'} \setminus S_n^{'} (\sigma_i)) \cup 
S_n^{(i-1)} (\sigma_i) \in \mathbb{S}_{n}$. 
Observe that if
$\min_{\sigma\in\mathcal{N}_{n}\left( \sigma_i \right)} \left|\sigma\right|
\geq\max_{\sigma\in S_n^{(i-1)}} \left|\sigma\right|,$ then
$\min_{\sigma\in S_{n}^{'} \left( \sigma_i \right)}
\left|\sigma\right|
\geq \max_{\sigma\in S_n^{(i-1)} (\sigma_i)} \left|\sigma\right|$
and $\min_{\sigma\in S_{n}^{'} \left( \sigma_i \right)}\left|\sigma\right|
\geq |\sigma_i| + 1$.
Therefore under the previous condition the code 
$(S_n^{'} \setminus S_n^{'} (\sigma_i)) \cup 
S_n^{(i-1)} (\sigma_i)$ would be a strictly better 
symmetric fix-free code than $S_n^{'}$ for any choice of source probabilities, 
contradicting the assumption that $S_n^{'} \in \mathbb{O}_{n}$.
Hence,
\begin{equation}
\min_{\sigma\in\mathcal{N}_{n}\left(\sigma_i \right)}\left|\sigma\right|<\max_{\sigma\in S_n^{(i-1)}}\left|\sigma\right|. \label{eq:L3}
\end{equation}
$S_n^{(i)}$ consists of the smallest n elements of
$S_n^{(i-1)} \cup \mathcal{N}_{n} \left( \sigma_i \right)
\setminus\left\{ \sigma_i \right\} $, so (\ref{eq:L3}) implies that
$\tilde{l}_n^{(i-1)} = \max_{\sigma\in S_n^{(i-1)}} |\sigma| \geq
\max_{\sigma\in S_n^{(i)}} |\sigma| = \tilde{l}_n^{(i)}.$
Hence, ${l}_n^{'} = \tilde{l}_n^{(h)} \leq \tilde{l}_n^{(0)} = {l}_n .$

To begin our argument for the remainder of Theorem~\ref{thm:sum}, 
recall our assumption that
\begin{displaymath}
S_n = S_n^{(0)} \overset{\sigma_1}\Rightarrow S_n^{(1)} 
\overset{\sigma_2}\Rightarrow S_n^{(2)} \overset{\sigma_3}\Rightarrow 
\dots \overset{\sigma_h}\Rightarrow S_n^{(h)} = S_n^{'}
\end{displaymath}
is a shortest sequence of codes in $\mathbb{S}_{n}$ transforming
$S_n$ to $S_n^{'}$ via uses of the $\Rightarrow$ operation.

Suppose $\{ \pi_{1,1} , \ \pi_{2,1} , \ \dots , \ \pi_{k,1} \}
= \{ \sigma_1 , \ \sigma_2 , \ \dots , \ \sigma_h \}  \; \cap \; S_n$
and the elements of \\ $\{ \sigma_1 , \ \sigma_2 , \ \dots , \ \sigma_h \} 
\setminus S_n$ each have a proper
prefix in the set $\{ \pi_{1,1} , \ \pi_{2,1} , \ \dots , \ \pi_{k,1} \}$.
Then each string $\sigma_{\iota} , \ \iota \in \{ 1 , \ \dots , \ h \}$,
can alternatively be labeled $\pi_{g,j}$, where
\begin{itemize} 
\item if $\sigma_{\iota} \in S_n$, then $j=1$ and 
$g= |\{ \sigma_1 , \ \sigma_2 , \ \dots , \ \sigma_{\iota} \} \cap S_n |$, and
\item if $\sigma_{\iota} \not\in S_n$, then $j$ is one more than 
the number of strings among
$\{ \sigma_1 , \ \sigma_2 , \ \dots , \ \sigma_{\iota} \} $ that have 
$\pi_{g,1}$ as a proper prefix.
\end{itemize} 
Let $\gamma_g$ be the number of strings, including $\pi_{g,1}$, among
$\{ \sigma_1 , \ \sigma_2 , \ \dots , \ \sigma_h \}$ which have
$\pi_{g,1}$ as a prefix.

Let $\rho_i , \ i \in \{ 1 , \ \dots , \ k \}$, be an arbitrary permutation
of $\{ 1 , \ \dots , \ k \}$.  Then Theorem~\ref{thm:Sigma} implies that if
$S_n^{'} \in \mathbb{O}_{n}$, we can study the transformation from $S_n$ to
$S_n^{'}$ through any ordering of $\{ \sigma_1 , \ \sigma_2 , \ \dots , 
\ \sigma_h \}$ of the form 
\begin{equation}
\{\pi_{\rho_1, 1}, \ \dots , \ \pi_{\rho_1, \gamma_{\rho_1}}, \ 
\pi_{\rho_2, 1}, \ \dots , \ \pi_{\rho_2, \gamma_{\rho_2}}, \ \dots , \
\pi_{\rho_k, 1}, \ \dots , \ \pi_{\rho_k, \gamma_{\rho_k}} \}.
\label{eq:pi}
\end{equation}

We will use induction on $k$ to show that the condition
$\sum_{j=1}^n l_j^{'} \; \geq \; \sum_{j=1}^n l_j $
implies that $S_n^{'} \not\in \mathbb{O}_{n}$. 
For the basis step, Proposition~\ref{prop:sum} treats the case $k=1.$ 
For the inductive step, we assume the result is true when $k\leq\kappa$
and show that it is consequently true at $k=\kappa+1.$

We will consider the possible transformations from $S_{n}$ to 
$S_{n}^{\prime}$ using a permutation of $\{ \sigma_1 , \ \sigma_2 , \ \dots , 
\ \sigma_h \}$ of the form (\ref{eq:pi}).  If $S_n^{'} \in \mathbb{O}_{n}$,
then by Theorem~\ref{thm:Sigma} we can define for 
$i \in \{ 1, \ \dots ,  \ \kappa + 1 \}$ a sequence of symmetric fix-free codes
$C_n^{(i,1)}, \ C_n^{(i,2)}, \ \dots , \ C_n^{(i,\gamma_i)} = I_n^{(i)} 
\in \mathbb{S}_{n}$ for which
\begin{displaymath}
S_n \overset{\pi_{i,1}}\Rightarrow C_n^{(i,1)} 
\overset{\pi_{i,2}}\Rightarrow \dots 
\overset{\pi_{i,\gamma_i}}\Rightarrow C_n^{(i,\gamma_i)} = I_n^{(i)}.
\end{displaymath}
For $1 \leq i \leq \kappa+1$,
let $(l_{1}^{\left(i\right)},\ldots,l_{n}^{\left(i\right)})$ denote the sorted
and non-decreasing sequence of codeword lengths of $I_n^{\left(i\right)}.$
If for any $i, \ \sum_{j=1}^{n}l_{j}\geq\sum_{j=1}^{n}l_{j}^{\left(i\right)},$
then the condition
$\sum_{j=1}^n l_j^{'} \; \geq \; \sum_{j=1}^n l_j $ implies that
$\sum_{j=1}^{n}l_{j}^{'} \geq\sum_{j=1}^{n}l_{j}^{\left(i\right)}$.
By the inductive hypothesis it follows from the transformation from
$I_n^{(i)}$ to $S_n^{'}$ that 
$S_{n}^{\prime}\notin\mathbb{O}_{n}.$

Therefore, assume for all $i\leq \kappa+1$ that 
$\sum_{j=1}^{n}l_{j}^{\left(i\right)}>\sum_{j=1}^{n}l_{j}.$
Define $\lambda_{i}$ as the smallest integer for which 
\begin{displaymath}
I_n^{\left(i\right)}\subseteq S_{n}\cup\mathcal{N}_{\lambda_{i}}^{*}
\left(\pi_{i,1}\right)\setminus \{ \left\{ \pi_{i,1}\right\} \cup \tilde{S}_{n}^{\lambda_{i}} \}.
\end{displaymath}
Let 
\begin{eqnarray*}
D_{i} & = & S_{n}\setminus I_n^{\left(i\right)} \\
D_{i}^{\prime} & = & I_n^{\left(i\right)}\setminus S_{n} \\
m_{i} & = & \left|D_{i}\right|=\left|D_{i}^{\prime}\right|. 
\end{eqnarray*}
Let $\left(d_{i,1},\ldots,d_{i,m_{i}}\right)$
and $\left(d_{i,1}^{\prime},\ldots,d_{i,m_{i}}^{\prime}\right)$ be
the sorted and non-decreasing sequences of codeword lengths of $D_{i}$
and $D_{i}^{\prime},$ respectively. Then by (\ref{eq:C}) and (\ref{eq:D})
we have 
\begin{equation}
\sum_{j=1}^{t}d_{i,j}^{\prime}\geq\sum_{j=1}^{t}d_{i,j}, \; 1 \leq t \leq m_i,
\label{eq:E}
\end{equation}
and we also have 
\begin{align}
\left|\pi_{i,1}\right|=d_{i,1}<\lambda_{i}+1\leq d_{i,2}\leq\ldots\leq d_{i,m_{i}}\label{eq:F}\\
d_{i,1}+1\leq d_{i,1}^{\prime}\leq d_{i,2}^{\prime}\leq\ldots\leq d_{i,m_{i}}^{\prime}\leq\lambda_{i}\label{eq:G}
\end{align}

Define $\mu$ as the smallest integer for which 
\[
S_{n}^{\prime}\subseteq S_{n}\cup\left[\bigcup_{i=1}^{\kappa+1}\mathcal{N}_{\mu}^{*}\left(\pi_{i,1}\right)\right]\setminus\left[\bigcup_{i=1}^{\kappa+1}\left\{ \pi_{i,1}\right\} \cup \tilde{S}_{n}^{\mu} \right].
\]
Observe that 
\begin{equation}
\mu\leq\min\left\{ \lambda_{1},\ldots,\lambda_{\kappa+1}\right\} .\label{eq:G-1}
\end{equation}

Let $m=\left|S_{n}\setminus S_{n}^{\prime}\right|$, 
and let $\left(\delta_{1},\ldots,\delta_{m}\right)$
and $\left(\delta_{1}^{\prime},\ldots,\delta_{m}^{\prime}\right)$
be the sorted and non-decreasing sequences of codeword lengths of
$S_{n}\setminus S_{n}^{\prime}$ and $S_{n}^{\prime}\setminus S_{n},$
respectively. 
If $S_n^{'}$ is a candidate for membership in $\mathbb{O}_{n}$ and we are 
studying part of a shortest transformation from $R_n$ to $S_n^{'}$, then
because $\delta_{1},\ldots,\delta_{\kappa+1}$ are the ordered
lengths of $\pi_{1,1},\ldots,\pi_{\kappa+1,1}$, it follows that
\begin{align}
\delta_{1}\leq\delta_{2}\leq\ldots\leq\delta_{\kappa+1}<\mu+1\leq\delta_{\kappa+2}\leq\ldots\leq\delta_{m}\label{eq:H}\\
\delta_{1}+1\leq\delta_{1}^{\prime}\leq\delta_{2}^{\prime}\leq\ldots\leq\delta_{m}^{\prime}\leq\mu\label{eq:I}
\end{align}

As in the proof of Proposition~\ref{prop:sum}, we can argue that
$S_{n}^{\prime} \not\in \mathbb{O}_{n}$ if 
$\sum_{i=1}^{t}\delta_{i}^{\prime}\geq\sum_{i=1}^{t}\delta_{i}$
for all $t\leq m.$ 
The condition $\sum_{i=1}^n l_i^{'} \geq 
\sum_{i=1}^n l_i$ here implies that 
$\sum_{i=1}^{m}\delta_{i}^{\prime}\geq\sum_{i=1}^{m}\delta_{i}.$
To establish the remaining $m-1$ inequalities we consider three cases:
\begin{enumerate}
\item $t=1:$ We know that $\delta_{1}^{\prime}\geq\delta_{1}+1.$
\item $2\leq t\leq\kappa+1:$ Starting from $t=1$ we will sequentially map
each $t\leq\kappa+1$ into a different ordered pair 
$\left({i}(t),{j}(t) \right)$ satisfying 
$\delta_{t}^{\prime}=d_{{i}(t),{j}(t)}^{\prime}$ as follows.
If there are multiple unchosen pairs $(i(t),j(t))$ which satisfy 
the equality then we select the one with minimum ${j}(t)$ 
and then, if necessary, minimum ${i}(t)$.
Let 
\begin{align*}
{\cal I}_t & =\left\{ i:\tau\rightarrow\left(i,j\right)\textrm{ for some }\tau\leq t\right\} \\
j_{t}\left(i\right) & =\left|\left\{ j:\tau\rightarrow\left(i,j\right)\textrm{ for some }\tau\leq t\right\} \right|
\end{align*}
Then 
\begin{align*}
\sum_{a=1}^{t}\delta_{a}^{\prime} =\sum_{i\in {\cal I}_t}\sum_{j=1}^{j_{t}\left(i\right)} d_{i,j}^{\prime} 
  \overset{(a)}{\geq}\sum_{i\in {\cal I}_t}\sum_{j=1}^{j_{t}\left(i\right)}d_{i,j}
 & \overset{(b)}{\geq}\sum_{i\in {\cal I}_t}\left(d_{i,1}+\sum_{j=2}^{j_{t}\left(i\right)}\left(\lambda_{i}+1\right)\right)\\
 & \overset{(c)}{\geq}\sum_{i\in {\cal I}_t}\left(d_{i,1}+\sum_{j=2}^{j_{t}\left(i\right)}\mu\right)\\
 & \overset{(d)}{\geq}\sum_{a=1}^{t}\delta_{a} .
\end{align*}
Here (a) follows from (\ref{eq:E}), (b) follows from (\ref{eq:F}),
(c) follows from (\ref{eq:G-1}) and (d) follows from (\ref{eq:H})
and the assumption that $t\leq\kappa+1.$ 
\item $\kappa+2\leq t\leq m-1:$ We are given $\sum_{i=1}^{m}\delta_{i}^{\prime}\geq\sum_{i=1}^{m}\delta_{i}$
or, equivalently, $\sum_{i=1}^{\kappa+1}\left(\delta_{i}^{\prime}-\delta_{i}\right)\geq\sum_{i=\kappa+2}^{m}\left(\delta_{i}-\delta_{i}^{\prime}\right).$
(\ref{eq:H}) and (\ref{eq:I}) imply that for $i\geq\kappa+2$, 
\[
\delta_{i}\geq\delta_{i}^{\prime}+1 .
\]
Hence for $t\geq\kappa+2$, 
\[
\sum_{i=1}^{t}\left(\delta_{i}^{\prime}-\delta_{i}\right)\geq\sum_{i=t+1}^{m}\left(\delta_{i}-\delta_{i}^{\prime}\right)\geq0.
\]
\end{enumerate}
Thus the condition $\sum_{i=1}^n l_i^{'} \geq 
\sum_{i=1}^n l_i$ here implies that $S_{n}^{\prime} \not\in \mathbb{O}_{n}$.
\end{IEEEproof}

Theorem \ref{thm:sum} shows conditions for which the $n$ inequalities of 
Lemma~\ref{lemma:dominant} can be reduced to one.  We next show that if by 
an application of Proposition~\ref{prop:sum} or Theorem~\ref{thm:sum}
we determine that $S_n^{'} \not\in \mathbb{O}_{n}$, then
we can automatically conclude that certain related codes also are not members
of $\mathbb{O}_{n}$.  We have the following result. 
\begin{theorem}
Suppose that the codes $S_n, \ S_n^{'}, \ C_n \in \mathbb{S}_{n}$,
that $S_n$ is in a shortest transformation from $R_n$ to $S_n^{'}$
through a sequence of $\Rightarrow$ operations, and 
that $S_n^{'}$ is in a shortest transformation from $R_n$ to $C_n$
through a sequence of $\Rightarrow$ operations.
Let $(l_1, \ l_2, \ \dots , \ l_n)$ and 
$(l_1^{'}, \ l_2^{'}, \ \dots , \ l_n^{'})$ be the sorted and non-decreasing 
sequences of codeword lengths of $S_n$ and $S_n^{'}$, respectively.
Suppose that $\sum_{j=1}^n l_j^{'} \; \geq \; \sum_{j=1}^n l_j $. 
If the portion of the shortest transformation from $S_n$
to $S_n^{'}$ satisfies either
\begin{itemize}
\item $S_n \overset{\pi_1}\Rightarrow S_n^{'}$ or 
\item there is a sequence of symmetric fix-free codes 
$S_{n}^{\left(1\right)}, \ S_{n}^{\left(2\right)}, \ \ldots , \
S_{n}^{\left(h \right)}\in\mathbb{S}_{n}$
for some $h \geq 2$ with
\[
S_{n}=S_{n}^{\left(0\right)} \overset{\pi_1}\Rightarrow 
S_{n}^{\left(1\right)} \overset{\pi_2}\Rightarrow S_{n}^{\left(2\right)} 
\overset{\pi_3}\Rightarrow
\cdots \overset{\pi_h}\Rightarrow S_{n}^{\left(h\right)}=S_{n}^{\prime}
\]
and with $\pi_1$ being a prefix of $\pi_i$ for $i \geq 2$,
\end{itemize}
and the portion of the shortest transformation from $S_n^{'}$
to $C_n$ can be described for some $\eta \geq 1$ by 
\begin{displaymath}
S_n^{'} \overset{\sigma_1}\Rightarrow C_n^{(1)} 
\overset{\sigma_2}\Rightarrow C_n^{(2)} \overset{\sigma_3}\Rightarrow 
\dots \overset{\sigma_{\eta}}\Rightarrow C_n^{(\eta)} = C_n 
\end{displaymath}
with $\pi_1$ not being a prefix of $\sigma_i$ for $1 \leq i \leq \eta$,
then $C_n \not\in \mathbb{O}_{n}$. 
\label{thm:extension}
\end{theorem}
\begin{IEEEproof}
Following the notation introduced in the proof of Proposition~\ref{prop:sum}, 
let 
\begin{eqnarray*}
D & = & S_{n}\setminus S_{n}^{\prime} \; = \; 
\{ \tilde{s}_1 , \ \dots , \ \tilde{s}_m \} ,
\\
D^{\prime} & = & S_{n}^{\prime}\setminus S_{n} \; = \; 
\{ {s}_1^{'} , \ \dots , \ {s}_m^{'} \} , \\
m & = & \left|D\right|=\left|D^{\prime}\right| ,
\end{eqnarray*} 
and let $\left(d_{1},\ldots,d_{m}\right)$
and $\left(d_{1}^{\prime},\ldots,d_{m}^{\prime}\right)$ 
respectively denote the sorted and non-decreasing sequences of codeword 
lengths of $D$ and $D^{\prime}.$  To arrive at a contradiction, suppose
$C_n \in \mathbb{O}_{n}$. Then by Lemma~\ref{lem:shortest}, $\pi_1$
must be a prefix of some element of $C_n$, and therefore
\begin{displaymath}
D^{'} \cap C_n \neq \emptyset .
\end{displaymath}
Suppose $|D^{'} \cap C_n| = k$.  By the definition of the $\Rightarrow$
operation, $D^{'} \cap C_n = \{ {s}_1^{'} , \ \dots , \ {s}_k^{'} \}$.

From the proof of Proposition~\ref{prop:sum} we saw that the condition
$\sum_{i=1}^{n} l_i^{'} \geq \sum_{i=1}^{n} l_i$ implies that
$\sum_{j=1}^{t} d_j^{'} \geq \sum_{j=1}^{t} d_j$ for all $1 \leq t \leq m$.
Therefore, the sequence of sorted and non-decreasing lengths of the
strings in $C_n \cup \{ \tilde{s}_1 , \ \dots , \ \tilde{s}_k \} \setminus
\{ {s}_1^{'} , \ \dots , \ {s}_k^{'} \}$ dominates the sequence of sorted
and non-decreasing codeword lengths of $C_n$.  To complete the proof it
suffices to show that $C_n \cup \{ \tilde{s}_1 , \ \dots , \ \tilde{s}_k \} 
\setminus \{ {s}_1^{'} , \ \dots , \ {s}_k^{'} \}  \in \mathbb{S}_{n}$. 
By the definition of the $\Rightarrow$ operation, we have that
$D \cap C_n = \emptyset$.  Furthermore, for $1 \leq i \leq \eta , \
\sigma_i \not\in D$ because either $\sigma_i \in S_n^{'}$
or $\sigma_i \in {\cal N}_n^{*} (\sigma_j)$ for some $j<i$ with
$\sigma_j \in S_n^{'}$.  Hence 
$C_n \cup \{ \tilde{s}_1 , \ \dots , \ \tilde{s}_k \} 
\setminus \{ {s}_1^{'} , \ \dots , \ {s}_k^{'} \}  \in \mathbb{S}_{n}$. 
\end{IEEEproof}

Recall that $R_n = \{s_1, \ s_2 , \ \dots , \ s_n \}$.
We have the following result.
\begin{cor}
Let $C^{\mbox{{\small prefix}}}$ be the set of palindromes (not including 1) 
which are proper prefixes of at least one codeword in $C_n \in \mathbb{O}_{n}$.
For $i \geq n/2, \ s_i \not\in C^{\mbox{{\small prefix}}}$.
\label{cor:half}
\end{cor}
\begin{IEEEproof}
For $i \geq (n+2)/2, \ \min_{\sigma \in {\cal N} (s_i)} |\sigma| = 2i-1
\geq n+1,$ so the $\overset{s_i}\Rightarrow$ operation would not produce a
code in $\mathbb{S}_{n}$. If $n$ is odd, then the shortest two palindromes
which have $s_{[(n+1)/2]}$ as a proper prefix have lengths $n$ and $n+1$.
If $n$ is even, then the shortest two palindromes which have $s_{[n/2]}$ 
as a proper prefix have lengths $n-1$ and $n$. In either of these cases
it is better to keep $s_{[n/2]}$ or $s_{[(n+1)/2]}$ as a codeword than to
turn it into a proper prefix of one.
\end{IEEEproof}

Observe that for a string $\sigma$ and its bitwise complement 
$\overline{\sigma}$, the lengths of strings in ${\cal N}_n (\sigma)$ 
will match those of their bitwise complements in 
${\cal N}_n (\overline{\sigma})$. Therefore, the previous result implies
that if $0 \in C^{\mbox{{\small prefix}}}$, then for $i \geq n/2, \; 
\overline{s_i} \not\in C^{\mbox{{\small prefix}}}$.
More generally if a code $S_n$ contains 
$\sigma$ and $\overline{\sigma}$, then one can impose an ordering
on them for $C^{\mbox{{\small prefix}}}$ and thereby reduce the number of 
strings to be considered for replacement at the next step.  Furthermore, 
we immediately obtain the following extension to Theorem~\ref{thm:extension}.
\begin{cor}
Suppose that the codes $S_n, \ S_n^{'} \in \mathbb{S}_{n}$ and that
$S_n$ is in a shortest transformation from $R_n$ to $S_n^{'}$
through a sequence of $\Rightarrow$ operations.
Let $(l_1, \ l_2, \ \dots , \ l_n)$ and 
$(l_1^{'}, \ l_2^{'}, \ \dots , \ l_n^{'})$ be the sorted and non-decreasing 
sequences of codeword lengths of $S_n$ and $S_n^{'}$, respectively.
Suppose that $\sum_{j=1}^n l_j^{'} \; \geq \; \sum_{j=1}^n l_j $. 
If the portion of the shortest transformation from $S_n$
to $S_n^{'}$ satisfies either
\begin{itemize}
\item $S_n \overset{\pi_1}\Rightarrow S_n^{'}$ or 
\item there is a sequence of symmetric fix-free codes 
$S_{n}^{\left(1\right)}, \ S_{n}^{\left(2\right)}, \ \ldots , \
S_{n}^{\left(h \right)}\in\mathbb{S}_{n}$
for some $h \geq 2$ with
\[
S_{n}=S_{n}^{\left(0\right)} \overset{\pi_1}\Rightarrow 
S_{n}^{\left(1\right)} \overset{\pi_2}\Rightarrow S_{n}^{\left(2\right)} 
\overset{\pi_3}\Rightarrow
\cdots \overset{\pi_h}\Rightarrow S_{n}^{\left(h\right)}=S_{n}^{\prime}
\]
and with ${\pi_1}$ being a prefix of ${\pi_i}$ for 
$i \geq 2$,
\end{itemize}
and if $\overline{\pi_1} \in S_n$,
then the code $S_n^{''}$ defined by
\begin{displaymath}
S_{n} \overset{\overline{\pi_1}}\Rightarrow \hat{S}_{n}^{\left(1\right)} 
\overset{\overline{\pi_2}}\Rightarrow \hat{S}_{n}^{\left(2\right)} 
\overset{\overline{\pi_3}}\Rightarrow \cdots 
\overset{\overline{\pi_h}}\Rightarrow \hat{S}_{n}^{\left(h\right)}
=S_{n}^{''}
 \end{displaymath}
is not an element of $\mathbb{O}_{n}$.  Furthermore, for $\eta \geq 1$ 
any code $\hat{C}_n$ related to $S_n^{''}$ by a transformation of the 
form
\begin{displaymath}
S_n^{''} \overset{\sigma_1}\Rightarrow \hat{C}_n^{(1)} 
\overset{\sigma_2}\Rightarrow \hat{C}_n^{(2)} \overset{\sigma_3}\Rightarrow 
\dots \overset{\sigma_{\eta}}\Rightarrow \hat{C}_n^{(\eta)} = \hat{C}_n 
\end{displaymath}
for which $\overline{\pi_1}$ not being a prefix of $\sigma_i$ for 
$1 \leq i \leq \eta$ satisfies $\hat{C}_n \not\in \mathbb{O}_{n}$. 
\label{cor:extension}
\end{cor}

There would be a further simplification in using these ideas to
generate all optimal symmetric fix-free codes if the following conjecture
holds:
\begin{conj}
Suppose that the codes $S_n, \ S_n^{'}, \ C_n \in \mathbb{S}_{n}$,
that $S_n$ is in a shortest transformation from $R_n$ to $S_n^{'}$
through a sequence of $\Rightarrow$ operations, and 
that $S_n^{'}$ is in a shortest transformation from $R_n$ to $C_n$
through a sequence of $\Rightarrow$ operations.
Let $(l_1, \ l_2, \ \dots , \ l_n)$ and 
$(l_1^{'}, \ l_2^{'}, \ \dots , \ l_n^{'})$ be the sorted and non-decreasing 
sequences of codeword lengths of $S_n$ and $S_n^{'}$, respectively.
Suppose that $\sum_{j=1}^n l_j^{'} \; \geq \; \sum_{j=1}^n l_j $. 
If for some $\eta \geq 1$ a shortest transformation from $S_n$
to $C_n$ can be described 
$S_n \overset{\pi}\Rightarrow S_n^{'} \overset{\sigma_1}\Rightarrow C_n^{(1)} 
\overset{\sigma_2}\Rightarrow C_n^{(2)} \overset{\sigma_3}\Rightarrow 
\dots \overset{\sigma_{\eta}}\Rightarrow C_n^{(\eta)} = C_n ,$
then $C_n \not\in \mathbb{O}_{n}$.  If in addition $\overline{\pi} \in S_n$
and $S_n \overset{\overline{\pi}}\Rightarrow S_n^{''}$, then
$S_n^{''}  \not\in \mathbb{O}_{n}$ and $S_n^{''}$ is not in any shortest
transformation from $R_n$ to a code in $\mathbb{O}_{n}$.
\label{conj}
\end{conj}

If this conjecture is true, then at each code $S_n$ generated as a candidate
member of $\mathbb{O}_{n}$ we need only consider additional transformations
involving codewords which when replaced will result in codes with smaller 
sums of codeword lengths than that of $S_n$. Furthermore we obtain constraints
on $C^{\mbox{{\small prefix}}}$ which may result in other reductions to our 
search space for optimal codes.  However, while this conjecture 
is open, one way to effectively use Theorems~\ref{thm:sum} and
\ref{thm:extension} is to establish for each code $S_n$ and string $\pi$
whether or not the conditions $S_n \overset{\pi}\Rightarrow S_n^{'}$ and
$\sum_{j=1}^n l_j^{'} \; \geq \; \sum_{j=1}^n l_j $  
imply that (1) $S_n$ also has a sum of codeword lengths which is at most that
of any code $C_n \in \mathbb{S}_{n}$
given by $S_n^{'} \overset{\sigma_1}\Rightarrow C_n^{(1)} 
\overset{\sigma_2}\Rightarrow C_n^{(2)} \overset{\sigma_3}\Rightarrow 
\dots \overset{\sigma_{\eta}}\Rightarrow C_n^{(\eta)} = C_n$, where 
$\pi$ is a prefix of $\sigma_i$ for each $1 \leq i \leq \eta$
or (2) the preceding sequence of code transformations is not associated with a
non-increasing sequence of maximum codeword lengths.
If these latter constraints can be verified for a given code $S_n$
and string $\pi$, then it can be concluded that $S_n^{'}$ is not in any 
shortest transformation from $R_n$ to any code in $\mathbb{O}_{n}$; as we
indicated earlier,
this places restrictions on $C^{\mbox{{\small prefix}}}$ for optimal codes.

We have mentioned earlier that $R_n \in \mathbb{O}_n$ for $n \geq 3$.
This is the only optimal symmetric fix-free code for $n=3$ and $n=4$.
We next describe some of the other codes in $\mathbb{D}_n$ for $n \geq 5$.
\begin{theorem}
Let $(l_1, \ l_2, \ \dots , \ l_n)$ be the sorted and non-decreasing 
sequence of codeword lengths for a code $S_n \in \mathbb{S}_n$ satisfying
$R_n \Rightarrow S_n$.  Then $S_n \in \mathbb{D}_n$ if 
$\sum_{i=1}^n l_i < {n(n+1)}/{2}.$ \label{thm:one}
\end{theorem}
\begin{IEEEproof}
Assume $j$ is the index for which $S_n \subseteq (R_n \setminus \{s_j\}) \cup
{\cal N}_n (s_j)$.
To arrive at a contradiction, suppose that there is a code
$C_n = \{ c_1, \ \dots , \ c_n \} \in \mathbb{S}_n$ which differs from 
both $S_n$ and its complementary code, satisfies 
$|c_1 | \leq |c_2 | \leq \dots \leq |c_n |$, and has the property that
\begin{equation}
\sum_{i=1}^{\kappa} |c_i | \leq \sum_{i=1}^{\kappa} l_i, \; \mbox{for all} \;
\kappa \in \{1, \ \dots , \ n\}. \label{eq:kappa}
\end{equation}
Since 
$\sum_{i=1}^n |c_i | < {n(n+1)}/{2}$, it follows that $C_n \neq R_n$.
By Lemma~\ref{lem:root}, each codeword of $C_n$ has a prefix in 
$R_n = \{s_1 , \ \dots , \ s_n \}$.  Since $C_n \neq R_n$, there 
exists $\iota$ and $\gamma$ such that $s_{\gamma}$ is a proper prefix of
$c_{\iota}$.  Let $k$ be the smallest index for which $s_k \not\in C_n$.
Therefore, since the shortest string of ${\cal N}_n (s_{\gamma} )$ 
has length $\max \{2 \gamma -1, \ 2\}$,
\begin{equation}
|c_{\iota} | \geq \max \{2 \gamma -1, \ 2\} \geq \max \{2k-1, \ 2\} .
\label{eq:kn}
\end{equation}
Since $C_n \in \mathbb{S}_n$ it follows that $2k-1 \leq n$.

We next show that the first $\max \{2k-2, \ 1\}$ sorted and non-decreasing
codeword lengths of $C_n$ satisfy
\begin{eqnarray}
|c_i | = i, & & i \leq k-1 \label{eq:cli} \\ 
|c_i | \geq i+1, & & k \leq i \leq \max \{2k-2, \ 1\}. \label{eq:clii} 
\end{eqnarray}
If $k=1$, then  $0, \ 1 \not\in C_n$, so $|c_1 | \geq 2$.
If $k \geq 2$, then (\ref{eq:cli}) holds because 
$\{s_1 , \ \dots , \ s_{k-1} \} \subset C_n$.
By the Kraft inequality, $|c_k | \geq k-1$ with strict inequality since
$(1, \ 2, \ \dots , \ k-1, \ k-1)$ is not a feasible sequence of codeword
lengths among symmetric fix-free codes. 
If $|c_k | = k$, then $\{s_1 , \ \dots , \ s_{k-1} \} \subset C_n$
implies $s_k \in C_n$, which contradicts the definition of $k$.
Therefore $|c_k | \geq k+1$.  For $k \geq 3, \; k+1 \leq i \leq 2k-2$, suppose 
that (\ref{eq:clii}) is not always true.  Then there is a smallest index 
$t \in \{k+1, \ \dots , \ 2k-2\}$ such that $|c_t | \leq t$.
Since $|c_{t-1} | \geq t$, we have
\begin{equation}
|c_{t-1} | = |c_t | = t \leq 2k-2. \label{eq:ctk}
\end{equation}
We next show that $c_t \in R_n$.  To arrive at a contradiction, suppose that
$s_q \in R_n$ is a proper prefix of $c_t$.  By the same argument as for
(\ref{eq:kn}), we have that $|c_t| \geq 2k-1$, which contradicts
(\ref{eq:ctk}).  Hence, $c_t \in R_n$.  The same argument implies that 
$c_{t-1} \in R_n$, but $|c_t | \neq |c_{t-1} |$ for different elements of 
$R_n$.  Thus, (\ref{eq:clii}) follows because (\ref{eq:ctk}) is false.

There are three cases to consider to establish the result:
\begin{enumerate}
\item $1 \leq k < j \leq n$:  Since $R_k$ is a subset of $S_n$, it follows
that $l_i \leq i$ for $i \leq k$.  Therefore, by (\ref{eq:cli})
and (\ref{eq:clii}), $\sum_{i=1}^k l_i < \sum_{i=1}^k |c_i |$, which 
contradicts (\ref{eq:kappa}).
\item $1 \leq k = j \leq n$:  We have $1, \ s_j \not\in C_n$, so each
codeword $c_i \in C_n$ has a prefix 
$w_i \in (R_n \setminus \{ s_j \}) \cup {\cal N}_n (s_j ).$
Furthermore, $S_n$ consists of the shortest $n$ strings in
$(R_n \setminus \{ s_j \}) \cup {\cal N}_n (s_j ).$
Therefore, for $i=1, \; |c_1| \geq |w_1| \geq l_1$, and so (\ref{eq:kappa})
implies that $c_1 = w_1$.  Next suppose that there is an index
$\lambda \in \{1, \ \dots , \ n-1\}$ such that $c_i = w_i$ for all 
$i \leq \lambda$.  Observe that if $w_{\lambda+1} \in
\{w_1 , \ \dots , \ w_{\lambda} \} = \{c_1 , \ \dots , \ c_{\lambda} \}$,
then $\{c_1 , \ \dots , \ c_{\lambda}, \ c_{\lambda+1} \}$, does not satisfy
the prefix condition and cannot be a symmetric fix-free code.  Hence,
$w_{\lambda+1} \not\in \{w_1 , \ \dots , \ w_{\lambda} \}$.
Therefore, $\{w_1 , \ \dots , \ w_{\lambda} , \ w_{\lambda+1} \}$ is a subset
of $(R_n \setminus \{ s_j \}) \cup {\cal N}_n (s_j )$ with 
$\lambda+1$ distinct elements.  Thus, 
$\sum_{i=1}^{\lambda+1} |c_i | \geq \sum_{i=1}^{\lambda+1} |w_i | \geq 
\sum_{i=1}^{\lambda+1} l_i$.
It follows from (\ref{eq:kappa}) that $\sum_{i=1}^{\lambda+1} |c_i | =
\sum_{i=1}^{\lambda+1} l_i$.  By induction $(|c_1 |, \ |c_2 |, \ \dots , \
|c_n |) = (l_1, \ \dots , \ l_n),$ which contradicts our earlier assumption.
\item $1 \leq j < k \leq n$:  Let $v$ be the shortest element of
${\cal N}_n (s_j )$.  Define the code 
$B_{2k-1} = (R_{2k-1} \setminus \{ s_j \} ) \cup \{ v \} 
\subseteq  (R_n \setminus \{ s_j \} ) \cup {\cal N}_n (s_j )$.  Let
$b_{sum}$ be the sum of lengths of the elements of $B_{2k-1}$.
Therefore,
\begin{eqnarray}
\sum_{i=1}^{2k-1} l_i & = & b_{sum} \label{eq:bsum} \\
& = & 2k^2 -k -j + \max \{ 2j-1 , \ 2 \} \nonumber \\
& \leq & 2k^2 -1 \label{eq:2k} \\
& = & \sum_{i=1}^{k-1} i + \left( \sum_{i=k}^{2k-2} (i+1) \right) + (2k-1) 
\nonumber \\
& \leq & \sum_{i=1}^{2k-1} |c_i |, \label{eq:b}
\end{eqnarray}
by (\ref{eq:cli}), (\ref{eq:clii}), and the fact that 
$|c_{2k-1} | \geq |c_{2k-2} | \geq 2k-1$.  The only way for (\ref{eq:b})
to be consistent with (\ref{eq:kappa}) is for (\ref{eq:bsum}),
(\ref{eq:2k}), and (\ref{eq:b}) all to be equalities.
In order for (\ref{eq:2k}) to be an equality, $j=1$ and $k=2$.
Since $j=1, \; S_n = \{00, \ 11, \ 010, \ 101, \ \dots \}$.
For (\ref{eq:b}) to be an equality, $|c_1 | = 1, \ |c_2 | = 3, \ |c_3 | = 3$.
Recall that we assume that $n \geq 5$.  Since $c_1 = 0, \
c_2, \ c_3 \in \{101, \ 111\}$, it follows that $|c_4 | \geq 4$.
However, for these choices of $S_n$ and $C_n$ we find that
$\sum_{i=1}^4 l_i = 10 < 11 \leq \sum_{i=1}^4 |c_i |$, which again
contradicts (\ref{eq:kappa}).
\end{enumerate}

Since each way of constructing the symmetric fix-free code $C_n$ results in
a violation of an assumption, we find that $S_n \in \mathbb{D}_n$.
\end{IEEEproof}

One can use the experimental results of \cite{stak} to show that $R_n$ and 
the optimal codes of Theorem~\ref{thm:one} make up all of the optimal
codes for $n \leq 10$.  

Our last technical result establishes a special case of Conjecture~\ref{conj}.
\begin{theorem}
Suppose symmetric fix-free codes $S_n^{'}$ and $C_n$ are related to each
other and to $R_n$ by $R_n \overset{s_{\iota}}\Rightarrow S_n^{'}
\overset{\sigma}\Rightarrow C_n$, and suppose $S_n^{'} \not\in \mathbb{D}_n$.
Then $C_n \not\in \mathbb{O}_n$.
\label{thm:two}
\end{theorem}
\begin{IEEEproof}
The case where $\sigma=s_{\gamma}$ for $\gamma \neq \iota$ is established by
Theorem~\ref{thm:extension}.
Therefore we will assume that $\sigma \in {\cal N}_n (s_{\iota})$.
As usual, let $l_n^{'}$ denote the maximum codeword length of $S_n^{'}$.
We will prove the result by arguing that
$\min_{\psi \in {\cal N} (\sigma)} | \psi | > l_n^{'}$.
Because of the structure of $s_{\iota}$, it is simple to establish that
$|\sigma| \geq 2 \iota -1$ and $\min_{\psi \in {\cal N} (\sigma)} | \psi | 
\geq 3 \iota -2$.  Therefore it suffices to show
\begin{equation}
3 \iota -2 > l_n^{'} .
\label{eq:2.1}
\end{equation}

As in earlier proofs, let
\begin{equation}
m = |S_n^{'} \setminus R_n| = |R_n \setminus S_n^{'}| .
\label{eq:2.2}
\end{equation}
Let $0^r$ denote a string of $r$ zeroes and let $\theta^{\rho}$ denote a
palindrome of length $\rho$.  The shortest elements of
${\cal N}_n (s_{\iota})$ are of the form $10^{\iota-2}10^{\iota-2}1, \ 
s_{\iota}s_{\iota} , \ s_{\iota} \theta^1 s_{\iota} , \ 
s_{\iota} \theta^2 s_{\iota} , \ \dots , \ 
s_{\iota} \theta^{\iota -3} s_{\iota} .$
For $0 \leq \rho \leq \iota -3$, every palindrome $\theta^{\rho}$ satisfies
$s_{\iota} \theta^{\rho} s_{\iota} \in {\cal N} (s_{\iota}).$ 
Since there are $2^{\lfloor (\rho + 1)/2 \rfloor}$ palindromes of length 
$\rho$, if $m \leq \sum_{t=0}^{\iota-2} 2^{\lfloor t/2 \rfloor}$, then
we have a complete description of $S_n^{'} \setminus R_n$.

In Corollary~\ref{cor:half} we showed the desired result when 
$\iota \geq n/2$. Therefore, we need only consider the case where
$\iota \leq (n-1)/2$.  Suppose that for some $k \geq 1$,
\begin{equation}
l_n^{'} = 2 \iota + k.
\label{eq:2.3}
\end{equation}
It follows from (\ref{eq:2.1}) and (\ref{eq:2.3}) that we would like to show
\begin{equation}
k \leq \iota - 3.
\label{eq:2.4}
\end{equation}
Note that the preceding condition would also imply that the longest codeword
of $S_n^{'} \setminus R_n$ is of the form $s_{\iota} \theta^{k} s_{\iota} $
for an arbitrary length-$k$ palindrome $\theta^{k}$
and that we could completely describe $S_n^{'} \setminus R_n$.

By the definition of the $\Rightarrow$ operation, 
$R_n \setminus S_n^{'} = \{ s_{\iota}, \ s_{2\iota+k+1}, \ s_{2\iota+k+2}, \
\dots , \ s_n \}$, and the sum of the lengths of these words is
\begin{equation}
\iota + \sum_{i=2\iota + k + 1}^n i = \iota + n(m-1) - \frac{(m-1)(m-2)}{2}.
\label{eq:2.5}
\end{equation}
In order to find $k$, we wish to have
\begin{equation}
| {\cal N}_{2 \iota + k-1} (s_{\iota}) | 
< m \leq | {\cal N}_{2 \iota + k} (s_{\iota}) | .
 \label{eq:2.6}
\end{equation}
If (\ref{eq:2.4}) holds and $k$ is odd, then $k$ satisfies
\begin{equation}
\sum_{t=0}^{k} 2^{\lfloor t/2 \rfloor}
= 2^{ (k+3)/2 } -2
< m \leq 2^{(k+3)/2 } + 2^{(k+1)/2 } -2
= \sum_{t=0}^{k+1} 2^{\lfloor t/2 \rfloor} .
 \label{eq:2.7}
\end{equation}
If (\ref{eq:2.4}) holds and $k$ is even, then $k$ satisfies
\begin{equation}
\sum_{t=0}^{k} 2^{\lfloor t/2 \rfloor}
= 2^{ (k+2)/2 } + 2^{ k/2 } -2
< m \leq  2^{ (k+4)/2 } -2
= \sum_{t=0}^{k+1} 2^{\lfloor t/2 \rfloor} .
 \label{eq:2.8}
\end{equation}

Observe that if (\ref{eq:2.4}) holds, then the sum of the codeword lengths
over the set $S_n^{'} \setminus R_n$ is
\begin{equation}
m(2 \iota -1) + \sum_{t=0}^{k} t \cdot 2^{\lfloor t/2 \rfloor}
+ (k+1) \left(m - \sum_{t=0}^{k} 2^{\lfloor t/2 \rfloor} \right)
= m(2 \iota +k) - \sum_{t=0}^{k} (k+1-t) \cdot 2^{\lfloor t/2 \rfloor} .
 \label{eq:2.9}
\end{equation}

Since $S_n^{'} \not\in \mathbb{D}_n$, we have that the sum of codeword lengths
over $R_n \setminus S_n^{'}$ is at most the sum of codeword lengths over
$S_n^{'} \setminus R_n$.  Hence (\ref{eq:2.9}) and (\ref{eq:2.5}) imply that
 \begin{equation}
\iota + n(m-1) - \frac{(m-1)(m-2)}{2} \leq 
m(2 \iota +k) - \sum_{t=0}^{k} (k+1-t) \cdot 2^{\lfloor t/2 \rfloor} .
 \label{eq:2.10}
\end{equation}
Since
 \begin{equation}
m = |R_n \setminus S_n^{'}| = 
| \{ s_{\iota}, \ s_{2\iota+k+1}, \ s_{2\iota+k+2}, \ \dots , \ s_n \} |
= n+1- 2\iota -k, 
 \label{eq:2.11}
\end{equation}
the condition (\ref{eq:2.10}) can be rewritten
 \begin{equation}
\frac{n}{2} \geq \frac{m^2 -k-1}{2} 
+ \sum_{t=0}^{k} (k+1-t) \cdot 2^{\lfloor t/2 \rfloor} .
 \label{eq:2.12}
\end{equation}

Because of (\ref{eq:2.11}), the condition (\ref{eq:2.4}) that we wish to 
establish is equivalent to
 \begin{equation}
\frac{n}{2} \geq \frac{m+3k+5}{2} .
 \label{eq:2.13}
\end{equation}
Therefore, to demonstrate (\ref{eq:2.13}) it sufficient to show that
\begin{displaymath}
\frac{m^2 -k-1}{2} 
+ \sum_{t=0}^{k} (k+1-t) \cdot 2^{\lfloor t/2 \rfloor} 
\geq \frac{m+3k+5}{2} 
\end{displaymath}
or 
 \begin{equation}
\frac{m^2 -m}{2} -2k -3 
+ \sum_{t=0}^{k} (k+1-t) \cdot 2^{\lfloor t/2 \rfloor} 
\geq 0.
 \label{eq:2.14}
\end{equation}

If $k$ is odd, then 
$\sum_{t=0}^{k} (k+1-t) \cdot 2^{\lfloor t/2 \rfloor}= 
7 \cdot 2^{(k+1)/2} -2k-9$,
and we wish to verify if 
 \begin{equation}
\frac{m^2 -m}{2} + 7 \cdot 2^{(k+1)/2} -4k -12 \geq 0 
 \label{eq:2.15}
\end{equation}
when $m$ satisfies (\ref{eq:2.7}). The expression $m^2-m$ is minimized when
$m= 2^{(k+3)/2} -1$, and for this $m$ the left-hand side of (\ref{eq:2.15})
is $2^{k+2}+4\cdot 2^{(k+1)/2}-4k-11 \geq 1$ for $k \geq 1$.
If $k \geq 2$ is even, one can show that
$\sum_{t=0}^{k} (k+1-t) \cdot 2^{\lfloor t/2 \rfloor}= 10 \cdot 2^{k/2} -2k-9$,
and we wish to assess if 
 \begin{equation}
\frac{m^2 -m}{2} + 10 \cdot 2^{k/2 } -4k -12 \geq 0 
 \label{eq:2.16}
\end{equation}
when $m$ satisfies (\ref{eq:2.8}). The expression $m^2-m$ is minimized when
$m=3 \cdot 2^{k/2} -1$, and for this $m$ the left-hand side of (\ref{eq:2.16})
is $9 \cdot 2^{k-1}+5.5 \cdot 2^{k/2}-4k-11 \geq 10$ for $k \geq 2$.

Since (\ref{eq:2.14}) holds for all $k \geq 1$, the result follows.
\end{IEEEproof}

\begin{figure}[h!]
  \caption{A directed tree illustrating $\mathbb{D}_{20}$  }
  \centering
\begin{tikzpicture}[node distance   = 1.5 cm]

  \tikzset{VertexStyle/.style = {draw, shape          = circle,
                                 text           = black,
                                 inner sep      = 2pt,
                                 outer sep      = 1pt,
                                 minimum size   = 24 pt}}
  \tikzset{EdgeStyle/.style   = {->,>=stealth',shorten >=1pt,auto,
			thick,
                                 double  ,
                                 double distance = 1pt}}
  \tikzset{LabelStyle/.style =   {
                                  fill           = white,
                                  text           = black,
				sloped}}
     \node[VertexStyle](1){210};

     \node[VertexStyle,below=2.5 cm of 1](2){130};
     \node[VertexStyle,right=of 2](3){136};
     \node[VertexStyle,right=of 3](4){145};
     \node[VertexStyle,right=of 4](5){161};
     \node[VertexStyle,right=of 5](6){177};
     \node[VertexStyle,right=of 6](7){191};
     \node[VertexStyle,right=of 7](8){202};

     \node[VertexStyle,below=2.5 cm of 2](9){113};
     \node[VertexStyle,right=of 9](10){119};
     \node[VertexStyle,right=of 10](11){127};
     \node[VertexStyle,right=of 11](12){128};
     \node[VertexStyle,right=of 12](13){131};
     \node[VertexStyle,right=of 13](14){135};
     \node[VertexStyle,right=of 14](15){143};

     \node[VertexStyle,below=2.5 cm of 9](16){106};
     \node[VertexStyle,right=of 16](17){111};
     \node[VertexStyle,right=of 17](18){112};
     \node[VertexStyle,right=of 18](19){116};
     \node[VertexStyle,right=of 19](20){126};

     \node[VertexStyle,below=2.5 cm of 17](21){110};

     \draw[EdgeStyle](1) to node[LabelStyle][right]{0} (2);
     \draw[EdgeStyle](1) to node[LabelStyle][right]{11} (3);
     \draw[EdgeStyle](1) to node[LabelStyle][right]{101} (4);
     \draw[EdgeStyle](1) to node[LabelStyle][right]{1001} (5);
     \draw[EdgeStyle](1) to node[LabelStyle][right]{10001} (6);
     \draw[EdgeStyle](1) to node[LabelStyle][right]{100001} (7);
     \draw[EdgeStyle](1) to node[LabelStyle][right]{1000001} (8);

     \draw[EdgeStyle](2) to node[LabelStyle][right]{11} (9);
     \draw[EdgeStyle](2) to node[LabelStyle][right]{101} (10);
     \draw[EdgeStyle](2) to node[LabelStyle][right]{1001} (11);
    
     \draw[EdgeStyle](3) to node[LabelStyle][right]{101} (12);
     \draw[EdgeStyle](3) to node[LabelStyle][right]{111} (13);
     \draw[EdgeStyle](3) to node[LabelStyle][right]{1001} (14);

     \draw[EdgeStyle](4) to node[LabelStyle][right]{1001} (15);

     \draw[EdgeStyle](9) to node[LabelStyle][right]{00} (16);
     \draw[EdgeStyle](9) to node[LabelStyle][right]{101} (17);
     \draw[EdgeStyle](9) to node[LabelStyle][right]{111} (18);

     \draw[EdgeStyle](10) to node[LabelStyle][right]{010} (19);

     \draw[EdgeStyle](12) to node[LabelStyle][right]{111} (20);

     \draw[EdgeStyle](17) to node[LabelStyle][right]{010} (21);
  \end{tikzpicture}
    \label{g_2:graph_20}
\end{figure}
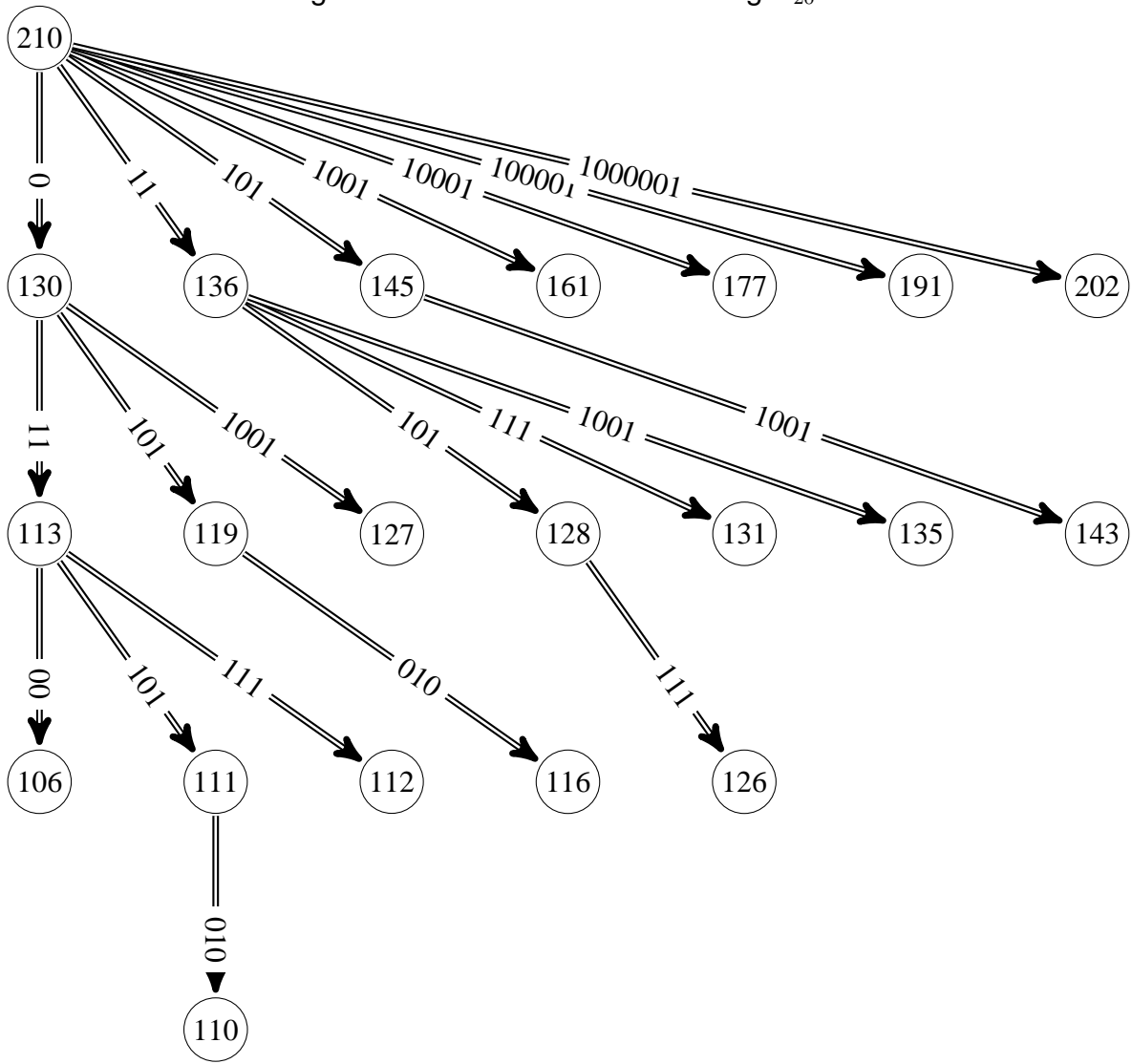

In Figure~\ref{g_2:graph_20}, we illustrate the tree of all $21$ codes in 
$\mathbb{D}_{20}$.
The numbers within the vertices represent the sum of codeword lengths for the 
corresponding code.  The strings labeling the edges
represent the codeword removed to go from a code to the next one.
The codelength sequences discussed in \cite{pr} form a lattice instead of a
tree.  Furthermore in \cite{pr} 
the codelength sequence with minimum sum was the furthest away from that
corresponding to the
most imbalanced code, while this is not the case here.  However, both here
and in \cite{pr} the most imbalanced (optimal) code of the class being 
studied had a central role in a mathematical analysis of optimal codes.


\paragraph*{Acknowledgment}  The authors were supported by NSF Grant 
No.~CCF-1017303.

\section*{Appendix}
Table~\ref{table:number} shows the exact number of 
different sorted and ascending codelength sequences for Huffman codes
(i.e., binary prefix condition codes which satisfy the Kraft inequality
with equality)
and an upper bound for the counterpart for optimal symmetric fix-free codes 
with $n$ words based on the number of dominant codelength sequences
when $n \leq 30$.  
The numbers for the Huffman code are taken from \cite{eng}, and the
numbers for dominant length sequences for symmetric fix-free codes 
come from \cite{aks, stak}.
\begin{table}
\caption{Number of (Sorted and Nondecreasing) Dominant Codelength Sequences 
over a Binary Code Alphabet}
\label{table:dominant}
\begin{center}
\begin{tabular}{|c|c|c|} \hline
$n$ & Huffman & Symmetric \\ \hline 
2 & 1 & 1 \\
3 & 1 & 1 \\
4 & 2 & 1 \\
5 & 3 & 2 \\
6 & 5 & 2 \\
7 & 9 & 3 \\
8 & 16 & 3 \\
9 & 28 & 4 \\
10 & 50 & 4 \\
11 & 89 & 6 \\
12 & 159 & 6 \\
13 & 285 & 8 \\
14 & 510 & 11 \\
15 & 914 & 11 \\
16 & 1639 & 13 \\
17 & 2938 & 13 \\
18 & 5269 & 17 \\ 
19 & 9451 & 18 \\ 
20 & 16952 & 21 \\ 
21 & 30410 & 22 \\ 
22 & 54555 & 24 \\ 
23 & 97871 & 26 \\ 
24 & 175588 & 29 \\ 
25 & 315016 & 32 \\ 
26 & 565168 & 34 \\ 
27 & 1013976 & 36 \\ 
28 & 1819198 & 42 \\ 
29 & 3263875 & 43 \\ 
30 & 5855833 & 46 \\ \hline
\end{tabular}
\end{center}
\label{table:number}
\end{table}

\end{document}